# High-Capacity Rechargeable Li/Cl$_2$ Batteries with Graphite Positive Electrodes

Guanzhou Zhu[1*], Peng Liang[1*], Cheng-Liang Huang[2,3], Cheng-Chia Huang[2], Yuan-Yao Li[2], Shu-Chi Wu[1], Jiachen Li[1], Feifei Wang[1], Xin Tian[1], Wei-Hsiang Huang[4,5], Shi-Kai Jiang[6], Wei-Hsuan Hung[7], Hui Chen[8], Meng-Chang Lin[8], Bing-Joe Hwang[6], Hongjie Dai[1]

[1]Department of Chemistry and Bio-X, Stanford University, Stanford, California 94305, USA.
[2]Department of Chemical Engineering, National Chung Cheng University, Chia-Yi 62102, Taiwan.
[3]Department of Electrical Engineering, National Chung Cheng University, Chia-Yi 62102, Taiwan.
[4]Graduate Institute of Applied Science and Technology, National Taiwan University of Science and Technology, Taipei 10607, Taiwan.
[5]National Synchrotron Radiation Research Center, Hsinchu 30076, Taiwan.
[6]Department of Chemical Engineering, National Taiwan University of Science and Technology, Taipei 10607, Taiwan.
[7]Institute of Materials Science and Engineering, National Central University, Taoyuan City 32001, Taiwan.
[8]College of Electrical Engineering and Automation, Shandong University of Science and Technology, Qingdao, Shandong Province, 266590, P. R. China.

*These authors contributed equally to this work.

**Abstract**

Developing new types of high-capacity and high-energy density rechargeable battery is important to future generations of consumer electronics, electric vehicles, and mass energy storage applications. Recently we reported ~ 3.5 V sodium/chlorine (Na/Cl$_2$) and lithium/chlorine (Li/Cl$_2$) batteries with up to 1200 mAh g$^{-1}$ reversible capacity, using either a Na or Li metal as the negative electrode, an amorphous carbon nanosphere (aCNS) as the positive electrode, and aluminum chloride (AlCl$_3$) dissolved in thionyl chloride (SOCl$_2$) with fluoride-based additives as the electrolyte[1]. The high surface area and large pore volume of aCNS in the positive electrode facilitated NaCl or LiCl deposition and trapping of Cl$_2$ for reversible NaCl/Cl$_2$ or LiCl/Cl$_2$ redox reactions and battery discharge/charge cycling. Here we report an initially low surface area/porosity graphite (DGr) material as the positive electrode in a Li/Cl$_2$ battery, attaining high battery performance after activation in carbon dioxide (CO$_2$) at 1000 °C (DGr_ac) with the first discharge capacity ~ 1910 mAh g$^{-1}$ and a cycling capacity up to 1200 mAh g$^{-1}$. Ex situ Raman spectroscopy and X-ray diffraction (XRD) revealed the evolution of graphite over battery cycling, including intercalation/de-intercalation and exfoliation that generated sufficient pores for hosting

LiCl/$Cl_2$ redox. This work opens up widely available, low-cost graphitic materials for high-capacity alkali metal/$Cl_2$ batteries. Lastly, we employed mass spectrometry to probe the $Cl_2$ trapped in the graphitic positive electrode, shedding light into the Li/$Cl_2$ battery operation.

**Introduction**

In recent decades there has been a steady increase in the demand of batteries for a wide range of applications from small personal electronics like mobile phones to medium ones like electric vehicles (EVs) and satellites, and to massive ones for grid scale energy storage. The developments of batteries with higher specific capacity, higher energy density, and longer cycle life have become increasingly important. Different types of batteries have thus been invented with various high-energy metals as anodes[2-10]. Recently we reported the discovery of rechargeable sodium/chlorine (Na/$Cl_2$) and lithium/chlorine (Li/$Cl_2$) batteries using either a Na or Li metal as the negative electrode, an amorphous carbon nanosphere (aCNS) as the positive electrode, and an electrolyte comprised of aluminum chloride ($AlCl_3$) and fluoride-based electrolyte additives dissolved in thionyl chloride ($SOCl_2$)[1]. The batteries operated based on the redox between either Na/$Na^+$ or Li/$Li^+$ at the negative electrode and $Cl^-$/$Cl_2$ at the positive electrode, delivering discharge voltage of ~ 3.5 V, cycling capacity of up to 1200 mAh $g^{-1}$ (based on aCNS mass) and cycle lives up to ~ 200 cycles[1]. Comparisons of several high surface area (~ 3000 $m^2$ $g^{-1}$) amorphous carbon positive electrode materials suggested the importance of large pore volume (~ 2.5 $cm^3$ $g^{-1}$), especially micropores, for effective trapping of $Cl_2$ to afford reversible $Cl^-$/$Cl_2$ redox[1]. Investigating various forms of carbon materials represents one of the main directions of alkali metal/$Cl_2$ batteries. Crystalline graphite materials are typically of low surface area and pore volume, and are seemingly unlikely candidates for high-capacity porous electrode materials needed for depositing large amount of metal chlorides and trapping and electro-reduction of chorine.

Herein, we investigate a defective graphite material (Micro850, Asbury Carbons, abbreviated as DGr, see Methods) as the positive electrode in a Li/$Cl_2$ battery. High performance battery was achieved using positive electrodes comprised of DGr activated in a carbon dioxide ($CO_2$) environment at 1000 ºC for 45 minutes (abbreviated as DGr_ac, see Methods). Despite the low surface area and pore volume of the as-made DGr_ac (SA < 20 $m^2$ $g^{-1}$, pore volume < 0.08

cm$^3$ g$^{-1}$), the resulting Li/Cl$_2$ battery delivered ~ 1910 mAh g$^{-1}$ first discharge capacity (based on the carbon mass). The battery was rechargeable and cyclable at a high specific capacity of up to 1200 mAh g$^{-1}$ with an average discharge voltage of ~ 3.5 V. We employed ex situ Raman spectroscopy, XRD and mass spectrometry to investigate the graphite evolution over repeated charge/discharge cycles. We found that the combination of CO$_2$ activation of disordered graphite and in situ electrochemical intercalation/deintercalation and exfoliation opened up the graphite structure for effective Cl$_2$ trapping, affording a high performance graphitic positive electrode for Li/Cl$_2$ battery.

**Result and Discussions**

The as received DGr material contained graphite flake with smooth edges and were several microns in size (see Fig. 1a scanning electron microscope (SEM) top image, also Fig. S1). After annealing in CO$_2$ at 1000 °C for 45 minutes (DGr_ac), we detected a substantial mass loss of ~ 68% and observed etch pits at the edges and in the middle of the planes of graphite flakes (Fig. 1a bottom image, Fig. S1, holes were indicated by arrows), suggesting the reaction CO$_2$ (g) + C (s) → 2 CO (g) initiated at defect sites in graphite[11]. Defects in DGr and DGr_ac were investigated by Raman spectroscopy (Fig. 1b). The relative intensity of the D band due to defects in the graphite, increased from ~ 0.05 in DGr (suggesting defects/disorder in the starting material) to ~ 0.08 in DGr_ac after CO$_2$ activation when normalized by the intensity of the G band arising from the doubly degenerate E$_{2g}$ phonon mode at the center of the Brillouin-zone (Fig. 1b)[12]. The intensity of the D' band originated from the intravalley scattering process by defects in the graphene plane also increased after DGr was treated in CO$_2$ (Fig. 1b)[13-15]. In addition, a slight blue shift in the G band position from ~1562.9 cm$^{-1}$ to ~ 1570.5 cm$^{-1}$ was detected after annealing in CO$_2$ (Fig. 1b), suggesting a decrease in the strain in the graphite lattice, which was also accompanied by a blue shift of the 2D band from ~ 2691.1 cm$^{-1}$ to ~ 2700.9 cm$^{-1}$ from DGr to DGr_ac (Fig. S2)[13, 16].

We constructed coin cell batteries by using a Li metal as the negative electrode, either DGr (Li/DGr battery) or DGr_ac (Li/DGr_ac battery) as the positive electrode in a neutral electrolyte (modified from our previous acidic electrolyte[1]) comprised of 1.8 M LiCl + 1.8 M AlCl$_3$ dissolved in SOCl$_2$ with 2 wt% of lithium bis(fluorosulfonyl)imide (LiFSI) added as the electrolyte (Fig. 1c, see Methods). The first discharge to 2 V (Fig. 1d) was due to the SOCl$_2$ reduction to S, SO$_2$ and with LiCl deposition on the positive electrode, delivering a voltage/capacity of ~ 3.18 V/~ 1391 mAh g$^{-1}$ and ~ 3.37 V/~ 1911 mAh g$^{-1}$ for the Li/DGr and Li/DGr_ac cells, respectively (Fig. 1d)[1, 17-21]. The improved first discharge capacity of DGr_ac over DGr was attributed to a ~ 42.9% increase in the surface area (13.11 m$^2$ g$^{-1}$ to 18.73 m$^2$ g$^{-1}$) and a ~ 40.0% increase in the pore volume (from 0.05 cm$^3$ g$^{-1}$ to 0.07 cm$^3$ g$^{-1}$) (Table S1), afforded by high temperature $CO_2$ activation that etched graphite at the edges and in-plane defects (Fig. 1a, Fig. S1). The larger surface area and pore volume provided more available sites to host the LiCl deposition accompanied by a higher first discharge capacity[1, 22]. In addition, a noticeable ~ 0.19 V increase in the first discharge voltage from ~ 3.18 V in the Li/DGr battery to ~ 3.37 V in the Li/DGr_ac battery was observed (Fig. 1d), suggesting possible catalytic effect for the formation and deposition of LiCl on DGr_ac on defects with oxygen containing functional groups resulted from $CO_2$ etching[23-24].

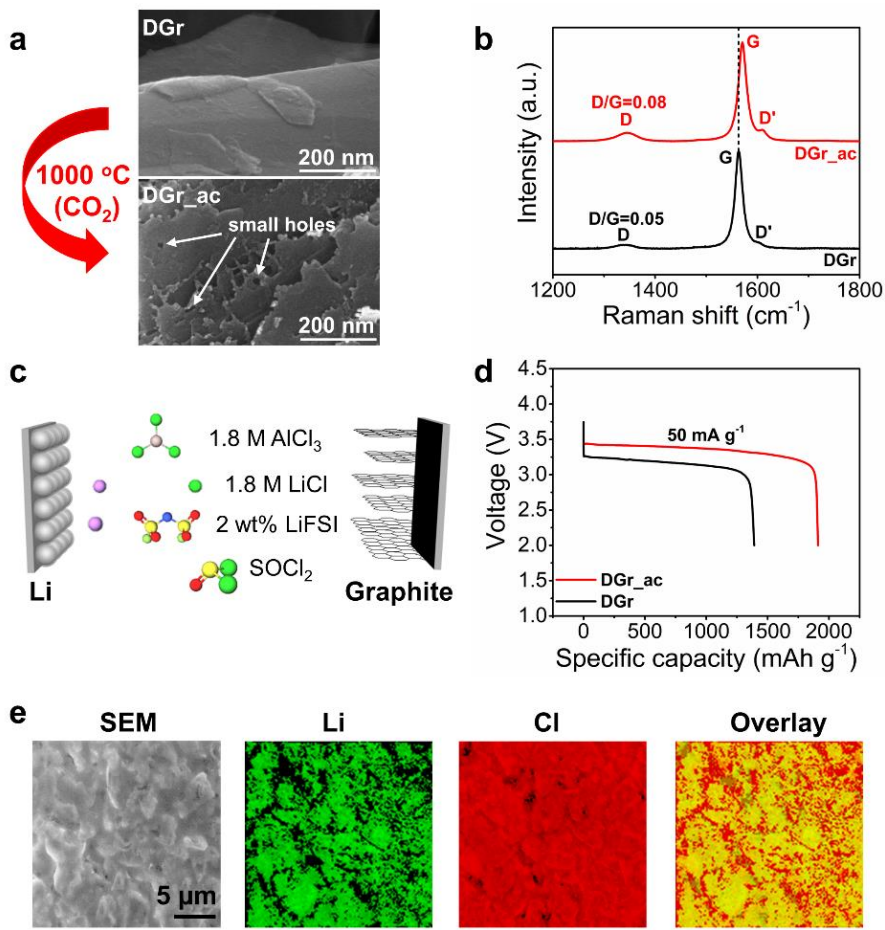

**Figure 1. A defective graphite (DGr) and CO$_2$ activated graphite (DGr_ac) for Li/graphite batteries in LiCl/AlCl$_3$/SOCl$_2$ electrolyte with 2 wt% of lithium bis(fluorosulfonyl)imide**

**(LiFSI) additive. a,** SEM images of as-received DGr and activated DGr_ac at the same magnification. $CO_2$ activation at 1000 °C led to more defects in DGr_ac at the edges and in the planes of the graphite flakes (indicated by arrows). **b,** Raman spectrum of as-received DGr and activated DGr_ac. The D/G ratio increased in DGr_ac and a slight blue shift was also observed in the DGr_ac spectrum, suggesting a decrease in the strain within the material. **c,** Schematic drawing of a Li/DGr or Li/DGr_ac battery. **d,** First discharge curves of Li/DGr and Li/DGr_ac batteries. Notice both the discharge capacity and discharge voltage increased when DGr_ac was the positive electrode. **e,** SEM image and elemental mappings of Li, Cl, and their overlay on a DGr_ac electrode after the first discharge. The electrode was covered by a LiCl layer. The mapping was done using an AES with a Scanning Auger Nanoprobe for imaging and elemental mapping.

We employed Auger Electron Spectroscopy (AES) with a Scanning Auger Nanoprobe for imaging and elemental mapping to examine the DGr_ac electrode after the first discharge, and observed a layer of LiCl covering the electrode resulted from the first discharge reaction (Fig. 1e). Such LiCl formation was also present in a DGr electrode after the first discharge, confirmed by XRD spectrum showing strong LiCl peaks (Fig. 2a).

The Li/DGr battery using graphite without $CO_2$ activation was rechargeable at 375 mAh g$^{-1}$ cycling capacity with the main charging plateau at ~ 4.00 V and another small plateau near the end of charging at ~ 4.06 V (Fig. 2b). Note that throughout this work, battery charging was controlled by setting the charging time depending on the cycling capacity (charging time = cycling capacity/current). The discharging step was controlled by setting a discharge cutoff voltage of 2 V. Two plateaus were also observed during discharging, with an initial small discharge plateau at ~ 3.69 V followed by the most dominant plateau at ~ 3.48 V (Fig. 2b). The main charging plateau at ~ 4.00 V was attributed to oxidation of LiCl to $Cl_2$, and the main discharging plateau at ~ 3.48 V was due to the reduction of $Cl_2$ back to LiCl[1, 19, 25]. The small plateau towards the end of charging at ~ 4.06 V was proposed to be the oxidation of electrolyte in forming $SCl_2$, $S_2Cl_2$, and $SO_2Cl_2$, and the small discharging plateau at ~ 3.69 V corresponded to the reduction of $SCl_2$ and $S_2Cl_2$[1, 19, 25].

At a capacity of 375 mAh g$^{-1}$, the Li/DGr battery could cycle stably over > 150 cycles (Fig. 2c). At this capacity, the battery's CE initially started at higher than 100% (Fig. 2c). Such high CE suggested 'in-situ activation' of DGr as the battery was cycling, i.e., through the battery electrochemical reactions especially during oxidative charging processes with the generation of Cl$_2$, the graphitic structure was continuously evolving and undergoing structural changes (see Fig. 4) such that after each charging step an increase in surface area/pore volume occurred, allowing additional SOCl$_2$ in the electrolyte to be reduced following the reduction of trapped Cl$_2$. This gave extra capacity towards the end of discharging (~ 3.15 V) and a CE > 100% (Fig. S3). The 'in-situ activation' of DGr eventually stopped once the CE of the battery stabilized at ~ 100% corresponding to reversible LiCl/Cl$_2$ redox (Fig. 2c). Note that the electrolyte additive LiFSI was found to prolong battery cycle life (Fig. 2d), attributed to a more stable solid-electrolyte interphase (SEI) on alkali metal negative electrode by introducing a fluoride component into the alkali metal chloride-rich SEI[1, 26-29].

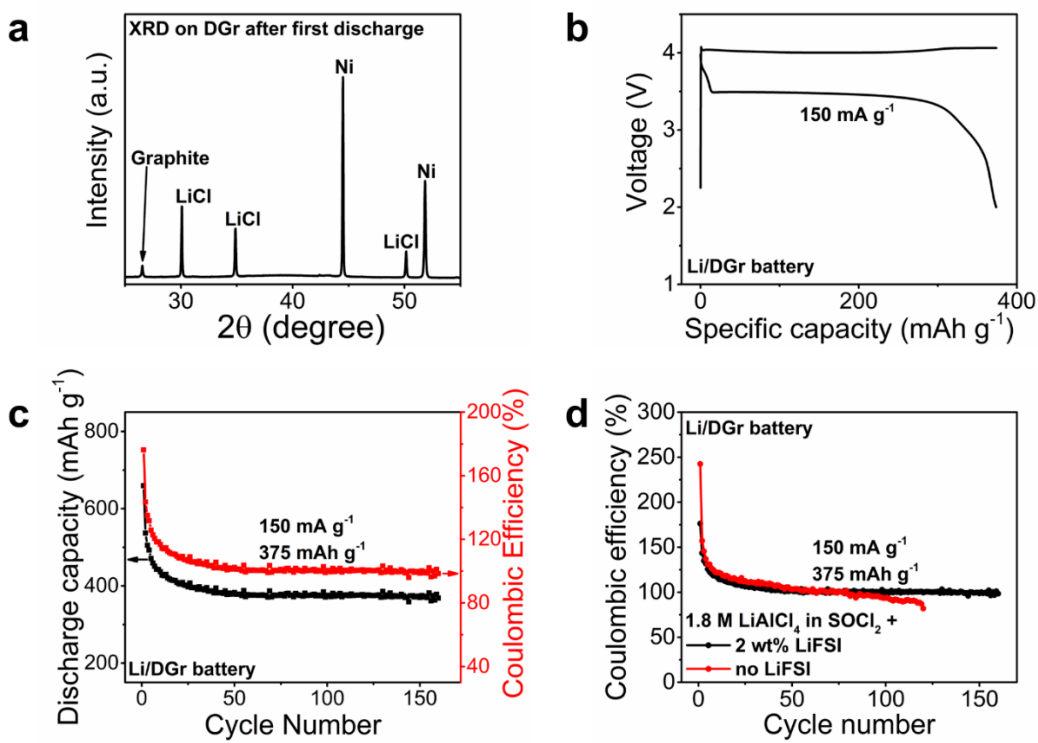

**Figure 2. Li/Cl$_2$ battery cycling performance with a Li/DGr cell using the as-received graphite as positive electrode without CO$_2$ activation. a,** XRD of DGr electrode after the battery's first discharge. LiCl was formed on the DGr electrode after first discharge, indicated by the strong LiCl XRD peaks. **b,** Typical charge-discharge curve of a Li/DGr battery cycling at 375 mAh g$^{-1}$ with 150 mA g$^{-1}$ current. The loading of DGr was ~ 4.3 mg cm$^{-2}$. The charging step was controlled by setting the charging time to be 2.5 hours at 150 mA g$^{-1}$ current (charging capacity = 375 mAh g$^{-1}$). The discharging step was controlled by setting a discharge cutoff voltage of 2 V. **c,** Cycling performance of a Li/DGr battery at 375 mAh g$^{-1}$ cycling capacity with 150 mA g$^{-1}$ current. The loading of DGr was ~ 4.3 mg cm$^{-2}$. **d,** Cycling performance comparison of Li/DGr batteries (375 mAh g$^{-1}$, 150 mA g$^{-1}$) with and without 2 wt% LiFSI added into the electrolyte. The battery with LiFSI as the electrolyte additive showed an improved cycling performance. The loading of DGr in both batteries was ~ 4.3 mg cm$^{-2}$.

The Li/DGr battery cycle life at higher capacities (e.g., 800 mAh g$^{-1}$) was much reduced (to < 50, Fig. S4a, b) over 375 mAh/g cycling. Importantly, activation of DGr in $CO_2$ at 1000 °C afforded Li/DGr_ac batteries with obviously improved cycling performance at higher capacities, with cycling life > 140 cycles and > 85 cycles at a cycling capacity of 600 mAh g$^{-1}$ and 800 mAh g$^{-1}$, respectively (Fig. 3a-c). As in the Li/DGr battery case, SEM imaging also revealed LiCl coating on DGr_ac positive electrode after discharge (Fig. 3d, Fig. S5 bottom row) and mostly disappeared after charging to 800 mAh g$^{-1}$ (Fig. 3e, Fig. S5 top row). Note that the total discharge capacity over > 140 cycles at 600 mAh g$^{-1}$ (Fig. 3a) was ~ 575 mAh, well exceeding the total theoretical capacity of ~ 103 mAh for reducing and consuming all the 150 μL $SOCl_2$ in the electrolyte. However, increasing the cycling capacity of the Li/DGr_ac battery to 1000 mAh g$^{-1}$ and 1200 mAh g$^{-1}$ led to a reduced cycle life of > 55 cycles and > 35 cycles, respectively (Fig. 3f, Fig. S6).

We consistently observed that the Li/$Cl_2$ batteries could be reversibly cycled with ~ 100% CE up to an upper limit of capacity (~ 375 mAh g$^{-1}$ for Li/DGr cells and ~ 1200 mAh g$^{-1}$ for Li/DGr_ac cells), with the limit depending on the type of carbon material used for the positive electrode or treatment method for activating the carbon. The first discharge of our batteries showed > 1300 mAh g$^{-1}$ capacity (Fig. 1d) through $SOCl_2$ reduction to LiCl deposited on the positive electrode. However, not all of the 1$^{st}$ discharge capacity was reversible/rechargeable, i.e., only a percentage of the 1$^{st}$ discharge deposited LiCl (residing in the small pores in the positive electrode) was reversibly oxidized to $Cl_2$ for subsequent reduction and battery cycling with a high CE ~ 100%, as we showed recently for the $Na/Cl_2$ cells[1]. In the case of Li/DGr battery, the reversible/rechargeable capacity was only ~ 375 mAh $g^{-1}$/~ 1391 mAh $g^{-1}$ (~ 27%). High temperature $CO_2$ activation of carbon positive electrode materials always led to an increase in the limit of reversible cycling capacity as in the current DGr_ac (up to ~ 63% of the 1$^{st}$ discharge capacity was reversible) vs. DGr case, attributed partly to increased surface area and pore volume. Another trend was that the battery cycle life increased at lower cycling capacity. When cycling/charging to higher capacities, higher degree of less reversible reactions (e.g., electrolyte oxidation) likely occurred near the end of the charging (evident by the slight increase in charging voltage, Fig. 3c) when most of the LiCl in the pores of carbon electrode was oxidized. It is an important finding here that carbon material choice/design and chemical/physical activation of carbon positive electrode materials could lead to much higher cycling capacity limits for $Li/Cl_2$ batteries with useful cycle lives.

We also investigated the rate capability of the Li/DGr_ac battery at 800 mAh $g^{-1}$ with currents varied from 50 mA $g^{-1}$ (0.0625 C) to 800 mA $g^{-1}$ (1C) (Fig. S7). The battery was cyclable at all these different current conditions but with its CE slightly decreased as the current was increased (CE = ~ 90% at 1 C, Fig. S7).

In a capacity retention study, we held an 800 mAh $g^{-1}$ charged Li/DGr_ac battery at open-circuit for 3 days and then discharged the battery, attaining ~ 100% CE without charge loss (Fig. 3g). Only a slight decrease in the discharge voltage from ~ 3.54 V to ~ 3.35 V was observed near the end of the discharge (Fig. 3h, red curve vs. black curve), attributed to the reduction of $SO_2Cl_2$ formed during the retention period by reactions between $SO_2$ and $Cl_2$ escaped from the trapping sites on carbon[1, 19, 25]. Such reaction was very slow without lowering battery CE and the battery was able to resume normal cycling after the long retention period (Fig. 3g, S8).

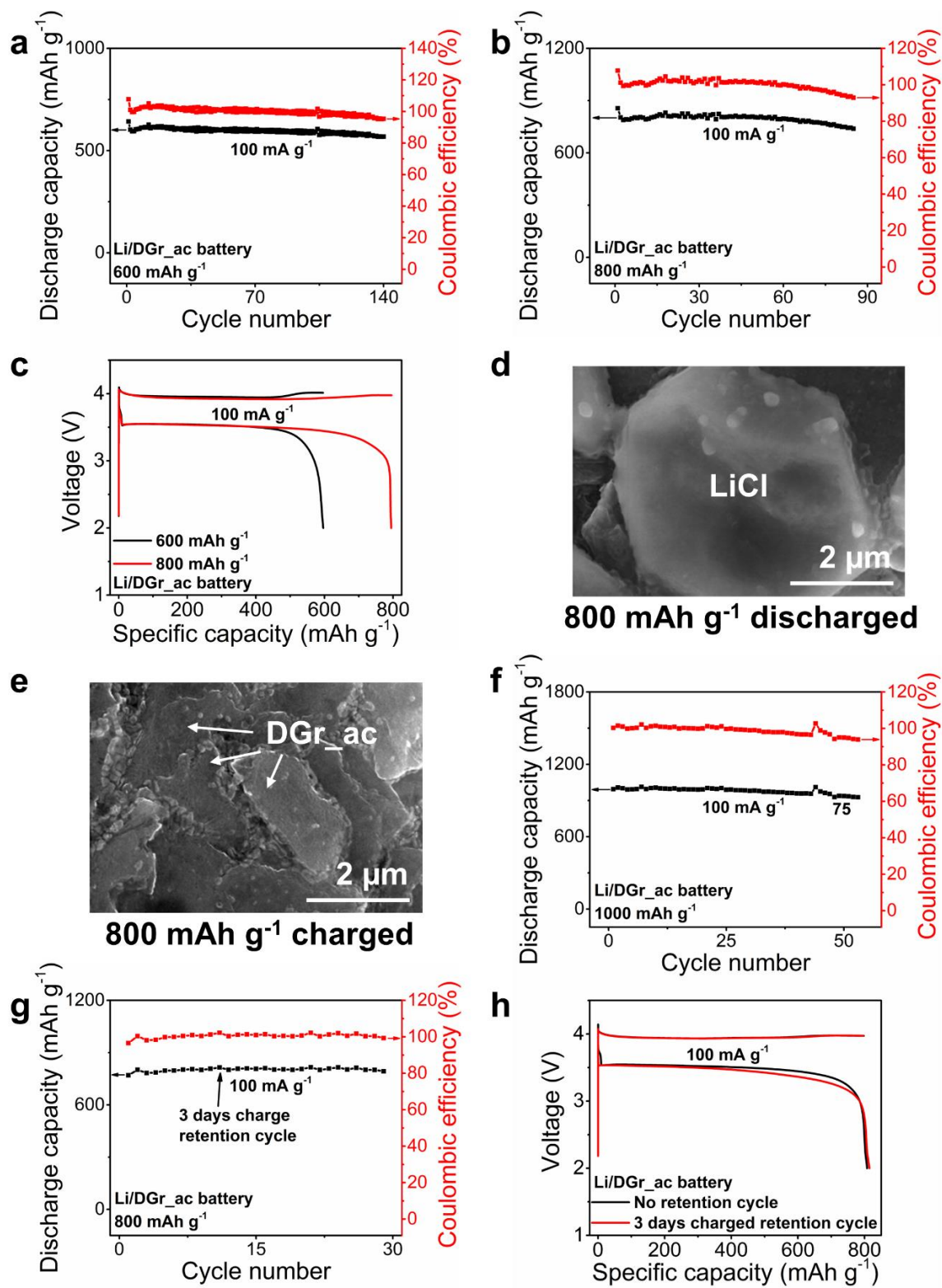

**Figure 3. Li/Cl₂ battery cycling performance with a Li/DGr_ac cell using CO₂ activated graphite as positive electrode. a,** Cycling performance of a Li/DGr_ac battery at 600 mAh g⁻¹

capacity with 100 mA g$^{-1}$ current. The battery was cyclable for more than 140 cycles. The loading of DGr_ac was ~ 3.8 mg cm$^{-2}$. Throughout this work, the charging step was controlled by setting the charging time depending on the cycling capacity (charging time = cycling capacity/current). The discharging step was controlled by setting a discharge cutoff voltage of 2 V. **b,** Cycling performance of a Li/DGr_ac battery at 800 mAh g$^{-1}$ capacity with 100 mA g$^{-1}$ current. The battery was cyclable for more than 85 cycles. The loading of DGr_ac was ~ 4.1 mg cm$^{-2}$. **c,** Typical charge-discharge curves of a Li/DGr_ac battery at 600 mAh g$^{-1}$ and 800 mAh g$^{-1}$ with 100 mA g$^{-1}$ current. **d-e,** SEM images of DGr_ac electrode after the battery was discharged (d) and charged (e) to 800 mAh g$^{-1}$. After charging, LiCl on the DGr_ac was oxidized/removed to form Cl$_2$ and the graphite flake underneath was exposed. After discharging, Cl$_2$ was reduced back to LiCl and passivated the DGr_ac electrode. **f,** Cycling performance of a Li/DGr_ac battery at 1000 mAh g$^{-1}$ with various currents (100 mA g$^{-1}$ and 75 mA g$^{-1}$). The battery was cyclable for more than 55 cycles. The loading of DGr_ac was ~ 3 mg cm$^{-2}$. **g,** Cycling performance of a Li/DGr_ac battery at 800 mAh g$^{-1}$ after holding the battery in open-circuit for 3 days in charged state in one cycle (indicated by arrow). The battery was able to stably cycle after the charged retention cycle. The loading of DGr_ac was ~ 2.6 mg cm$^{-2}$. **h,** Charge-discharge curve of the battery in a normal cycle versus in a cycle where the battery was held in open-circuit for 3 days after charging. The loading of DGr_ac was ~ 2.6 mg cm$^{-2}$.

The up to 1200 mAh g$^{-1}$ cycling capacity of Li/DGr_ac battery suggested high surface area/pore volume in the positive DGr_ac electrode for LiCl deposition and Cl$_2$ trapping, well exceeding those in the DGr electrode owing to CO$_2$ activation. The high capacity approached those of Li/Cl$_2$ and Na/Cl$_2$ cells using high surface area (~ 3000 m$^2$ g$^{-1}$) and large pore volume (~ 2.5 cm$^3$ g$^{-1}$) amorphous carbon nanospheres as positive electrode[1]. To investigate the evolution of DGr_ac through battery cycling, we first performed Raman spectroscopy at different stages of battery cycling (Fig. 4a). After the battery's first discharge, a blue shift in the graphite G band from ~ 1571.2 cm$^{-1}$ to ~ 1580.8 cm$^{-1}$ was observed, attributed to DGr_ac interactions with the oxidizing SOCl$_2$ in the electrolyte causing a Raman blue-shift due to hole doping (Fig. 4a, black curve vs. red and blue curves)[30]. Over cycling, the G band position remained constant, and upon battery charging (both in cycle 13 and cycle 40), a new peak at ~ 1600 cm$^{-1}$ appeared (Fig. 4a, labeled by '*'), suggesting graphite intercalation causing G band splitting most likely by AlCl$_{3.3}^{-1}$ species between the layers of positive charged/oxidized graphite facilitated by Cl$_2$ generated by LiCl electro-oxidation[2, 31-33]. The similar intensity of the ~ 1600 cm$^{-1}$ peak to the 1580 cm$^{-1}$ peak suggested an intercalation stage number $n$ = ~ 4 intercalation based on the equation $\frac{I_{\sim 1580}}{I_{\sim 1600}} = \sim \frac{n-2}{2}$ [32, 34]. Upon battery discharge the peak at 1600 cm$^{-1}$ mostly disappeared, indicating reversible intercalation/de-intercalation over charge/discharge cycling (Fig. 4a).

To eliminate the doping effects of the electrolyte on the Raman spectra of graphite, we washed DGr_ac electrodes using deionized ultra-filtered (DIUF) water to remove SOCl$_2$ and salts from the electrolyte. Raman measurements showed that all the charged and discharged DGr_ac electrodes at cycle 13 and cycle 40 displayed only 3 peaks (D band, G band, and D' band, similar to the starting DGr_ac, Fig. 4b) without any intercalation peak at 1600 cm$^{-1}$ due to reactions with water. The blue shift in the G band also disappeared after water washing (Fig. 4a, b), reversing the electrolyte hole-doping effect. An obvious increase in the D/G ratio was observed in the cycled DGr_ac after washing with DIUF water (Fig. 4b), suggesting that more defects were formed in the DGr_ac over cycling. The crystalline domain size L$_a$ of DGr_ac decreased over battery cycling from analysis of the D/G ratio inversely proportional to L$_a$[35-36]. We further analyzed the Raman intensity ratio between the 2D band (~ 2700 cm$^{-1}$) and the G band (~ 1570 cm$^{-1}$) (I$_{2D}$/I$_G$), observing an increase from ~ 0.29 in as made/CO$_2$-activated DGr_ac to an average of ~ 0.34 in DIUF-washed cycled DGr_ac (Fig. S9). I$_{2D}$/I$_G$ was known to be inversely proportional to the number of graphene layers in the graphite flake[37-38], indicating that as the Li/DGr_ac battery was cycled, a degree of exfoliations of graphite occurred to result in the reduced average number of graphene layers in DGr_ac.

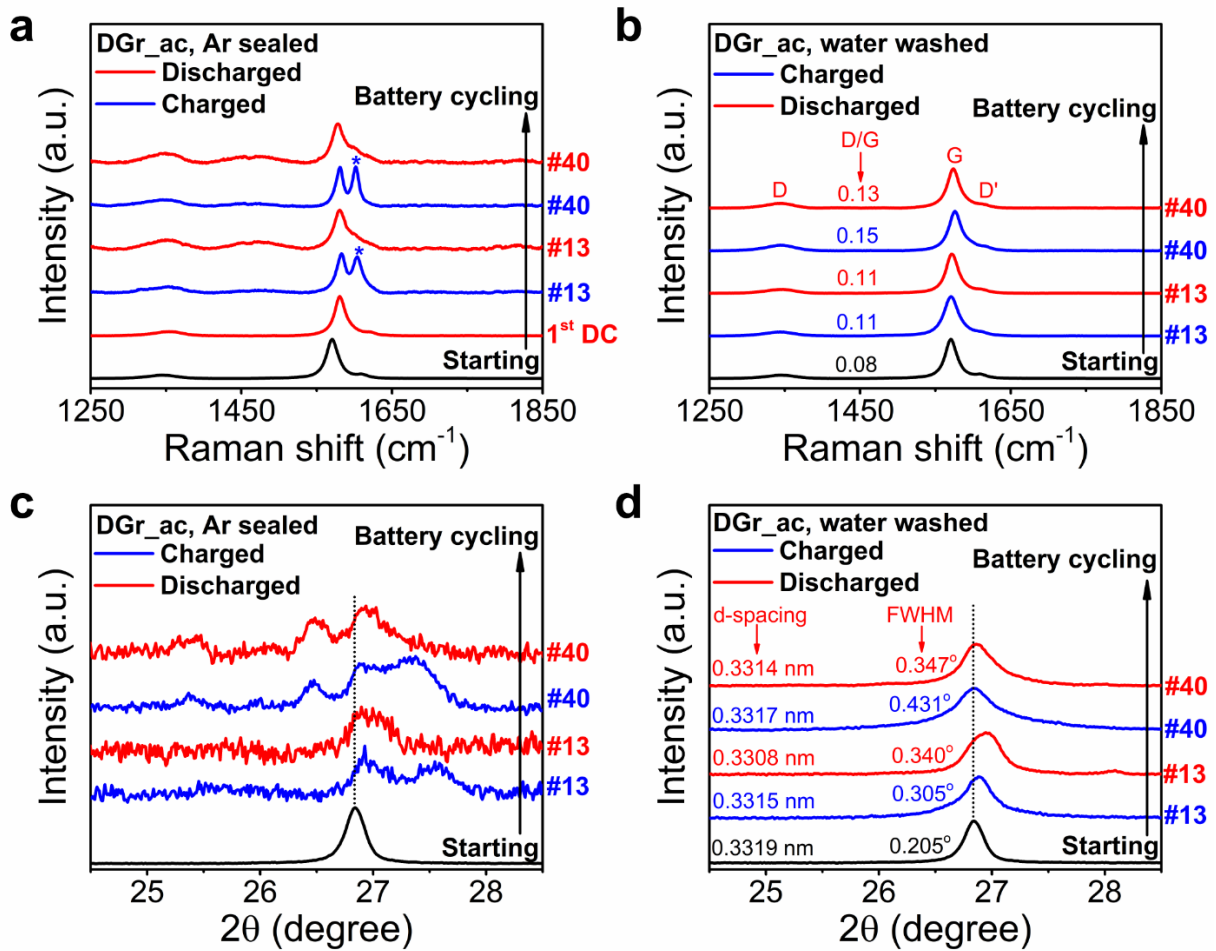

**Figure 4. Raman and XRD studies of DGr_ac positive electrodes at different stages over Li/Cl$_2$ battery cycling. a,** Raman spectra of DGr_ac in different states. Black, blue, and red curve indicated starting DGr_ac, charged DGr_ac, and discharged DGr_ac, respectively. The label on the right represented the cycle number of the Li/DGr_ac battery (1$^{st}$ DC = after first discharge). The G band peak splitting and reversal suggested intercalations and de-intercalations of DGr_ac during charging and discharging, respectively. **b,** Raman spectra of DGr_ac electrode at different states during battery cycling after washing the DGr_ac using water. The number labeled within the figure next to each spectrum was the intensity ratio between the D and G band of that spectrum. The label on the right represented the cycle number of the Li/DGr_ac battery. **c,** XRD spectra of starting DGr_ac (black curve) and charged (blue curve) and discharged (red curve) DGr_ac in cycle 13 and cycle 40. The label on the right represented the cycle number of the Li/DGr_ac battery. **d,** XRD spectra of DGr_ac at different states during battery cycling after washing the DGr_ac using DIUF water. The full width at half maximum (FWHM) of each peak and the corresponding

d-spacing were labeled next to each spectrum. The label on the right represented the cycle number of the Li/DGr_ac battery. All the batteries were cycled at 800 mAh g$^{-1}$ until they reached the designated states (first discharge, cycle 13, cycle 40). All batteries had similar DGr_ac loading at ~ 3.5 mg cm$^{-2}$.

To further glean the structural evolution of graphite, we performed XRD of DGr_ac over battery cycling ex situ (Figure 4c, d, see Methods). The cycled DGr_ac (in a special sample holder without exposing to air after removal from battery cells, see Methods) had its main XRD peak shifted to a slightly higher angle of ~ 2θ = 26.95° from 2θ = 26.84° for starting DGr_ac (Fig. 4c), and the peak was observed in all spectra regardless of whether the DGr_ac was in the charged or discharged state (Fig. 4c). On charging (13$^{th}$ and 40$^{th}$ charged state), a new peak at higher angle of ~ 2θ = 27.54° appeared in the XRD spectrum (Fig. 4c), suggesting compression of interplane d-spacing due to intercalation. In the charged state the lack of a well-defined diffraction peak at a smaller angle could correspond to interlayer expansion causing exfoliation with nanometer range gaps opened up between graphite sheets. The interlayer compression peak at ~ 2θ = 27.54° disappeared upon subsequent discharging, indicating reversibility (Fig. 4c). When a Li/DGr_ac battery was cycled to its 40$^{th}$ cycle, a new peak at a lower angle of ~ 2θ = 26.48° appeared, and was present in both the charged and discharged DGr_ac spectra (Fig. 4c). This result suggested expansion of interlayer spacing in the DGr_ac during the stable cycling phase of the battery, resulted from repeated intercalations/de-intercalations. At cycle 40, XRD peak at the higher angle (compression region) also appeared in charged DGr_ac and disappeared in the discharged DGr_ac, similar to those at cycle 13 (Fig. 4c), indicating reversibility of intercalation through cycling.

Upon expose to air and washing with DIUF water, XRD of the post-cycling DGr_ac electrodes showed only one main XRD peak with similar d-spacing to the as-activated DGr_ac (Fig. 4d). Importantly however, we found that the full width at half maximum (FWHM) of the XRD peak in the cycled DGr_ac increased over that in as-activated DGr_ac especially at cycle 40 (Fig. 4d, #13 versus #40), consistent with the exfoliation of graphite over cycling[39-41].

The Raman and XRD results revealed structural evolution of the graphite positive electrode over battery cycling in the electrolyte containing $SOCl_2$, $AlCl_4^-$, $Li^+$ and F-additives, as well as oxidative species such as $SO_2$, $SO_2Cl_2$ and $Cl_2$ resulted from battery cycling[1]. Repeated intercalation/deintercalation led to increased defects/disorder in the graphite and induced a degree of exfoliation, which could be responsible for opening up more pore-like sites for LiCl deposition and $Cl_2$ trapping for reversible $LiCl/Cl_2$ redox. It is important to note that a simple calculation showed that intercalation of graphite by chlorine species alone could not support a cycling capacity of > 800 mAh $g^{-1}$, since a 800 mAh $g^{-1}$ capacity would require a stage 1 intercalation with every carbon atom in graphite associating with ~ 0.18 associated $Cl_2$ (compared to the theoretical capacity of Li intercalated graphite with ~ 372 mAh $g^{-1}$ capacity in a stage 1 intercalation compound, with every carbon having ~ 0.17 associated $Li^+$)[42]. In our charged Li/DGr_ac battery, the intercalation stage was about 4, far from stage 1. Therefore, the reversible cycling capacities of 800-1200 mAh $g^{-1}$ were attributed to sufficient pore volumes opened up in the exfoliated graphite by repeated battery charging. These pores allowed filling of nanoscopic LiCl deposition upon discharge and trapping of the generated $Cl_2$ upon charging for subsequent chlorine reduction in battery discharge. The nature of the pores could be space between graphite sheets with nanometer scale gaps caused by exfoliation through battery cycling.

The starting graphite material DGr was several microns in size with pre-existing defects evident from Raman data of D/G = ~ 0.05 (Fig. 1b). Cycling of the Li/DGr battery led to gradual exfoliation and increased pores to allow additional $SOCl_2$ reduction for extra discharge capacity, signaled by the observed > 100 % CE before stabilizing at ~ 100% CE to support a ~ 375 mAh $g^{-1}$ reversible cycling capacity (Fig. 2c, S3). Treating DGr in $CO_2$ at 1000 °C led to increase defects and obvious etching at step edges and at defect sites in the plane, causing an obviously increase in the Raman D/G ratio (Fig. 1b). Such $CO_2$ 'activation' of graphite afforded an improved DGr_ac positive electrode material for our battery, likely by providing more weakened lattice sites for intercalation and exfoliation of graphite, opening up the graphite structure further with increased defect sites to facilitate $LiCl/Cl_2$ deposition/trapping and $Cl^-/Cl_2$ redox.

Lastly, to confirm $Cl_2$ trapping involved in the reversible $LiCl/Cl_2$ redox in the positive electrode, we employed mass spectrometry to analyze trapped species inside charged and discharged DGr_ac electrode (Fig. S10, see Methods). With a charged DGr_ac electrode (removed from a cycling battery at 800 mAh $g^{-1}$) under vacuum pumping, the detected $Cl_2$ pressure (m/z = 70 amu) normalized to $SOCl_2$ pressure (m/z = 118 amu) ($I_{Cl_2}/I_{SOCl_2}$) remained constant within the first ~ 10 hours of pumping (Fig. 5a blue curve), attributed to the removal of mostly electrolyte residing in the electrode with the detected $Cl_2$ arising from fragmentation of $SOCl_2$ in the electrolyte. After ~ 10 hours of pumping, $I_{Cl_2}/I_{SOCl_2}$ started to increase (Fig. 5a blue curve, Fig. 5b) due to the escaping of trapped $Cl_2$ (not from $SOCl_2$ fragments) in the pores of the DGr_ac and continued over time. In stark contrast, the same measurement performed with a discharged DGr_ac electrode at 800 mAh $g^{-1}$ found a constant $I_{Cl_2}/I_{SOCl_2}$ ratio throughout the entire time period, without trapped $Cl_2$ in a discharged DGr_ac (Fig. 5a red curve, Fig. 5c). In addition, we also heated charged and discharged DGr_ac electrodes at 80 °C for 2 hours and measured a much more rapidly increasing $I_{Cl_2}/I_{SOCl_2}$ ratio for the charged electrode over time (Fig. 5d blue curve, 5e). This suggested that heating facilitated more rapid escape of trapped $Cl_2$ from the charged DGr_ac electrode. On the other hand, heating a discharged DGr_ac electrode showed both the $I_{Cl_2}/I_{SOCl_2}$ ratio and the $Cl_2$ peak intensity in a $SOCl_2$-normalized mass spectrum remained nearly constant through heating (Fig. 5d red curve, 5f). Note that in a control experiment, we observed that a fresh electrolyte (1.8 M LiCl + 1.8 M $AlCl_3$ in $SOCl_2$ + 2 wt% LiFSI) showed similar mass spectrometry data as the discharged DGr_ac electrode (Fig. S11). Our mass spectrometry data suggested that the trapped $Cl_2$ in the DGr_ac electrode was stable for ~ 10 hours at room temperature under vacuum pumping, in line with the up to 3 days stability at room temperature inside a coin cell based on battery retention data (Fig. 3g, h). Taken together, the results confirmed that during battery charging, $Cl_2$ formed from the oxidation of LiCl was trapped in the pores of the electrode, which could then be reduced back to LiCl during subsequent discharging. This was a key to rechargeable $Li/Cl_2$ battery.

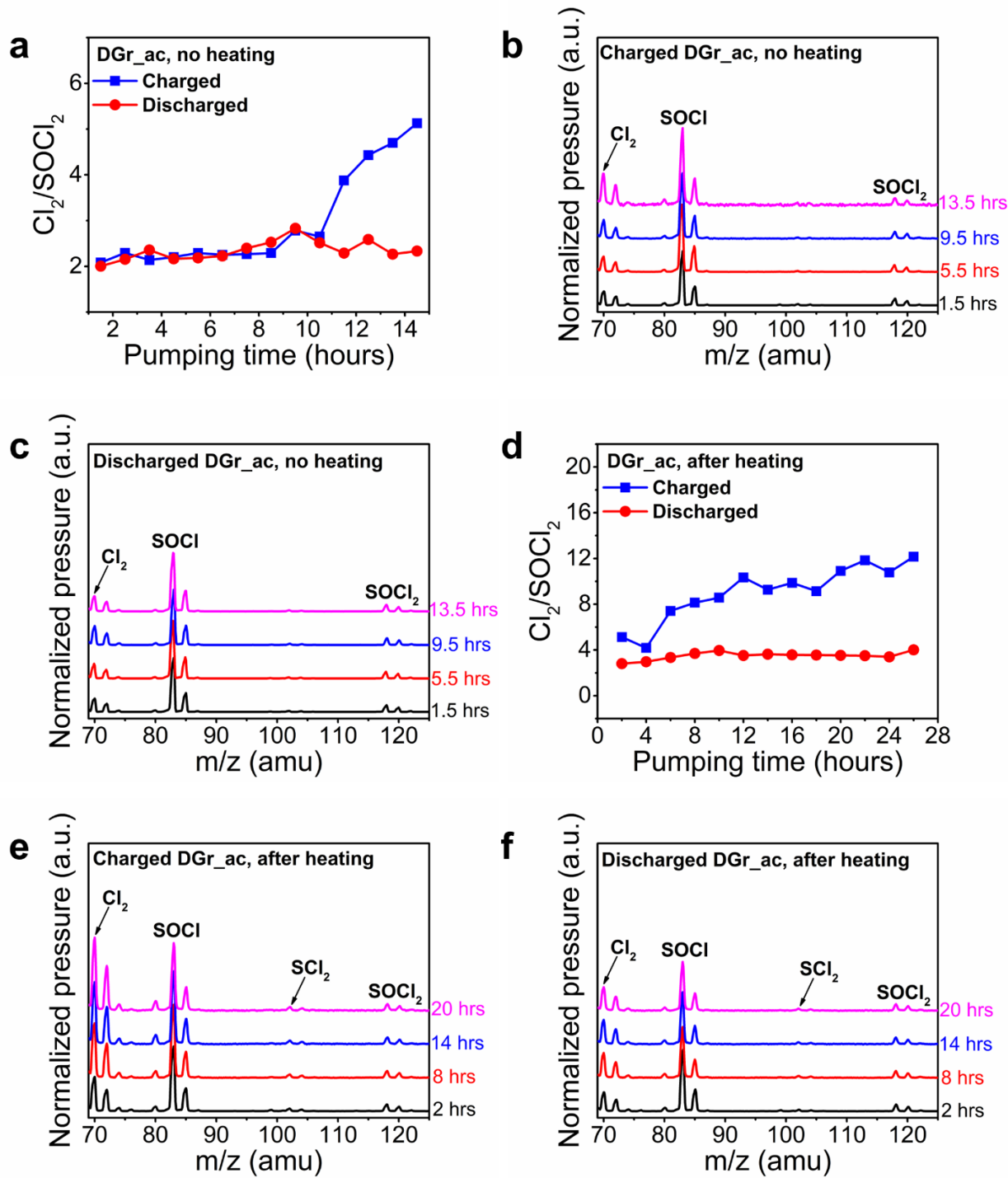

**Figure 5. Mass spectrometry of species released from charged and discharged DGr_ac electrodes in Li/DGr_ac batteries. a,** Under vacuum pumping, the detected $Cl_2$ (m/z = 70 amu) ratio to $SOCl_2$ (m/z = 118 amu) from a charged DGr_ac electrode versus from a discharged DGr_ac electrode over pumping time at room temperature. This ratio gradually increased after ~ 10 hours

in the charged electrode and remained nearly constant in the discharged electrode, suggesting the release of free $Cl_2$ trapped in the charged electrode but not from discharged electrode ($Cl_2$ detected in the discharged electrode case was from $SOCl_2$ fragmentation of electrolyte in the electrode). **b,** Mass spectra normalized by $SOCl_2$ molecular peak of a charged DGr_ac electrode recorded at several pumping times (indicated next to the spectrum). The normalized intensity of the $Cl_2$ peak clearly increased at longer pumping times. **c,** $SOCl_2$-normalized mass spectra of a discharged DGr_ac electrode at several pumping times (indicated next to the spectrum) showing constant $Cl_2$ peak over pumping (from fragmentation of $SOCl_2$ in the electrolyte). **d,** The detected ratio between $Cl_2$ (m/z = 70 amu) and $SOCl_2$ (m/z = 118 amu) in a charged DGr_ac electrode versus in a discharged DGr_ac electrode at different pumping times after these electrodes were heated at 80 ºC for 2 hours. The higher and much-faster increasing ratio in the charged DGr_ac electrode suggested that heating allowed more $Cl_2$ to escape from the charged electrode. This ratio remained nearly constant in a discharged DGr_ac electrode due to a lack of excess $Cl_2$ present in the sample. **e,** $SOCl_2$-normalized mass spectrum of a charged DGr_ac electrode after heating at 80 ºC for 2 hours at different pumping times (indicated next to the spectrum). The intensity of the $Cl_2$ peak became much stronger and increased more rapidly as pumping time increased. **f,** $SOCl_2$-normalized mass spectrum of a discharged DGr_ac electrode after heating at 80 ºC for 2 hours at different pumping times (indicated next to the spectrum). The intensity of the $Cl_2$ peak remained nearly constant in all the measured spectra. The batteries were charged and discharged at 800 mAh $g^{-1}$ with similar DGr_ac loading at ~ 2.5 mg $cm^{-2}$.

**Conclusion**

In this work, we found that defective microns scale graphite flakes could be used as the positive electrode in Li/$Cl_2$ batteries, especially after further 'activation' by annealing in a $CO_2$ environment at 1000 ºC to generate more defects and afford increased surface area and pore volume. The resulting Li/$Cl_2$ battery using DGr_ac positive electrode was rechargeable/cyclable at a capacity up to 1200 mAh $g^{-1}$ with an average discharging voltage of ~ 3.5 V. Investigations by Raman spectroscopy and XRD revealed the structural changes of the DGr_ac over batteries cycling caused by repeated intercalation/deintercalation and exfoliation, opening up pore-like spaces between graphite sheets for LiCl/$Cl_2$ trapping and reversible redox at high capacities. Mass spectrometry was employed to probe $Cl_2$ trapped in the charged graphite electrode, shedding light in battery operation. The results suggested that seemingly low surface area/small pore volume graphitic carbon materials are strong contenders for high performance Li/$Cl_2$ batteries, in addition to highly porous amorphous carbon materials.

**Methods**

**Activation/Annealing DGr in $CO_2$**

Micro850 from Asbury Carbons (DGr) was purchased and used directly. ~ 3 g of DGr was placed into an ignition dish and annealed using a horizontal tube furnace in a $CO_2$-flowing environment (flow rate 200 cc min$^{-1}$). The annealing temperature was 1000 $^o$C and the annealing time was 45 minutes. The temperature was increased from room temperature to 1000 $^o$C at a rate of 5 $^o$C min$^{-1}$. After annealing, the system was allowed to cool down naturally. ~ 68% mass loss was recorded after the annealing with the remaining graphite being DGr_ac.

**Fabrication of DGr and DGr_ac electrode**

90% by weight of either DGr or DGr_ac powder and 10% by weight of PTFE (60% aqueous dispersion, FuelCellStore) were mixed in 100% ethanol (Fisher Scientific). The mixture was then sonicated for 2 hours. Nickel (Ni) foam substrates were cut into circular pieces with diameter of 1.5 cm using a compact precision disk cutter (MTI, MSK-T-07). The circular shaped Ni foams were sonicated in 100% ethanol for 15 minutes and then dried at 80 $^o$C oven until all the ethanol evaporated. The weight of each circular shaped Ni foam was measured and recorded. The Ni foams were then hovered over a hot plate, and the mixture of DGr or DGr_ac, PTFE, and ethanol was then slowly dropped (180 µL each time) onto these circular-shaped Ni foams. The solvent from the previous drop must be completely dried before another drop was added to the Ni foam. The process was stopped when the desired amount of graphite was loaded onto each Ni foam. The graphite-loaded Ni foams were then dried at 80 $^o$C overnight and were then pressed using a spaghetti roller. The final weight of the graphite-loaded Ni foam was measured and recorded. The weight of graphite was determined by the final weight of graphite-loaded Ni foam minus the initial weight of the Ni foam times 90%. The graphite-loaded Ni foam was then ready to be used as the positive electrode in an actual battery.

**Electrolyte making**

Electrolyte was made inside an argon-filled glovebox. Thionyl chloride ($SOCl_2$) was purchased from Spectrum Chemical Mfg. Corp. (TH138-100ML) and used without any further purification. The appropriate amount of $SOCl_2$ was added into a 20-mL scintillation vial (Fisher Scientific). 1.8 M of aluminum chloride ($AlCl_3$, Fluka, 99%, anhydrous, granular) was weighed and added to the $SOCl_2$. The mixture was then stirred until all the $AlCl_3$ was dissolved. 1.8 M lithium chloride (LiCl, GanfengLithium) was then added into the solution and stirred for ~ 15 minutes until LiCl could not be further dissolved. In this step, we typically added a little excess of LiCl into the solution to make sure that the electrolyte was completely neutralized. 2 weight percent (2 wt%) of lithium bis(fluorosulfonyl)imide (LiFSI, Tianfu Chemical) was then added into the solution and stirred for ~ 45 minutes or until LiFSI could no longer be dissolved. Normally after electrolyte making, a small amount of residual salts (LiCl, LiFSI) remained in the solution. We then removed the vial from the stir plate to let it sit until all the residual salts sank to the bottom of the vial. The upper transparent layer of liquid was then ready to be used as the electrolyte.

**Battery making and testing**

All batteries were made inside an argon-filled glovebox. Lithium (Li) metal was purchased from Sigma-Aldrich. A nail file was used to scratch the Li metal to remove any surface contamination. The scratched Li metal was then pasted onto a coin cell spacer (MTI Corporation) and ready for use as the negative electrode. Either the DGr or DGr_ac electrode was put in the center of a positive CR2032 coin cell case (SS316, MTI corporation). 2 layers of quartz fiber filters (QR-100, Sterlitech) were put on top of the positive electrode as the separators. 150 µL of electrolyte was added onto the separators. The Li negative electrode was then put on top of the separators, with the Li metal directly facing the positive electrode. A piece of coin cell spring (MTI Corporation) was put on top of the spacer. Lastly, the negative CR2032 electrode case (SS316, MTI corporation) was put on top of the spring and the entire battery was sealed using a digital pressure controlled electric crimper for CR2032 coin cells with the digital pressure reading set to 13.2 (MTI corporation, MSK-160E). After the coin cell was assembled, the battery was transferred out of the glovebox. A layer of GE advanced silicone sealant was applied on the coin cell to cover the O-ring that sealed the two cases together. The purpose of this silicone layer was to prevent water and air from leaking into the battery. After the silicone was cured, the battery was tested using a battery tester (Neware, CT-4008-5V50mA-164-U). The charging step in battery testing was controlled by setting the charging time depending on the cycling capacity (mAh g$^{-1}$) and current condition (mA g$^{-1}$) (charging time = cycling capacity/current). The discharging step was controlled by setting a discharge cutoff voltage of 2 V.

**Raman spectroscopy measurements**

After the DGr_ac electrode has cycled to the designated state in a battery, the battery was disassembled and the DGr_ac electrode was taken out from the battery. A small hole was punched on an aluminum laminated pouch (MTI, EQ-alf-100-210) and the hole was covered by a small quartz window. The edge of the quartz window was sealed using GE advanced silicone sealant and the pouch was allowed to sit in air until the silicone was cured. A piece of carbon tape (Ted Pella, 16073) was pasted onto the pouch directly under the quartz window. The pouch was then transferred inside an argon-filled glovebox, and the DGr_ac electrode was pasted on the carbon tape so that it was visible via the quartz window. All sides of the pouch were then heat sealed inside the glovebox using a tabletop impulse sealer (Uline). After sealing, the pouch was transferred out the glovebox for Raman measurement using the Horiba Jobin Yvon (Olympus BX41) instrument with Ar$^+$ laser of 532 nm. Before each measurement, a piece of p-type boron doped silicon wafer was used for calibration and the silicon peak was adjusted to 520.7 cm$^{-1}$. After the instrument was calibrated, the Raman laser was focused on the DGr_ac sample through the quartz window and the Raman spectra were recorded.

Raman spectra on DIUF-water-washed samples were done similarly. The DGr_ac electrode was put inside a 20-mL scintillation vial and DIUF water was added to the vial to wash the sample. After washing, the electrode was air dried and its Raman spectra were recorded. No aluminum laminated pouch was needed for DIUF-water-washed samples as they were already exposed to air during washing.

**XRD measurement**

XRD measurement was conducted in the environmental measurements facility (Stanford Earth) using Rigaku MiniFlex 600 Benchtop X-ray Diffraction System. After the DGr_ac electrode has cycled to the designated state in a battery, the battery was disassembled and the DGr_ac electrode was taken out from the battery in an argon-filled glovebox. The DGr_ac was sealed inside a special XRD holder manufactured by Rigaku that consisted of a polycarbonate dome sitting on top of an O-ringed 1-inch holder with zero XRD background (Si (311)) in an argon-filled glovebox without exposure to air. The sample holder was then taken to the facility and placed inside the XRD instrument, and the measurements were taken at a scan rate of $2^o$ min$^{-1}$.

DIUF-water-washed samples were prepared similarly as those used for Raman measurements (see Raman spectroscopy measurements section above). The DIUF-water-washed samples were directly placed into the XRD instrument and didn't need the special XRD holder. The scan rate of the XRD measurement was $2^o$ min$^{-1}$.

**Scanning electron microscopy**

SEM measurements were conducted using a field-emission scanning electron microscopy (Hitachi S-4800). To characterize as-received DGr and activated DGr_ac, the samples were directly pasted on the SEM sample stage using double-sided carbon adhesive. To characterize the electrode inside a battery, the battery was first dissembled inside an argon-filled glovebox and the electrode was taken out from the battery. The electrode was firmly pasted on the SEM sample stage using double-sided carbon adhesive inside the glovebox. The sample stage was then sealed inside a Ziploc bag inside the glovebox and was immediately transferred out of the glovebox and put into the SEM instrument. During this transfer process, it was important to minimize the sample's exposure to air. The samples were observed using a 15 kV acceleration voltage and 10 µA emission current.

**Mass spectrometry measurements using residual gas analyzer**

Mass spectrometry measurements were conducted using a residual gas analyzer (Fig. S10, RGA-300, Stanford Research Systems). After the battery was cycled to its designated state, the battery and the sample chamber with valve #1 connected were both transferred inside an argon-filled glovebox (the sample chamber was disconnected from the RGA-300 setup from the disconnection point, Fig. S10). The battery was then disassembled and the DGr_ac electrode was taken out from the battery and immediately transferred into the sample chamber with valve #1 closed (Fig. S10). The sample chamber was then transferred outside the glovebox and connected back to the RGA-300 setup at the disconnection point (Fig. S10). Valve #2 was then opened with valve #1 remained closed for 1 hour so the turbo pump was pumping residual air in the instrument (Fig. S10). After 1 hour, valve #1 was opened, and species inside the DGr_ac electrode were continuously pumped to the RGA-300 detector via the capillary tube (Fig. S10). The mass spectrometry data was then recorded at different pumping time.

To measure the mass spectrometry data of a heated DGr_ac electrode, the DGr_ac electrode was first placed inside the sample chamber in an argon-filled glovebox (similar to above). Then with valve #1 closed (Fig. S10), the entire chamber was placed inside an 80 ºC oven for 2 hours. After heating, the chamber was transferred out from the oven and connected to the instrument at the 'disconnection point' (Fig. S10). With valve #1 closed, valve #2 was opened for 1 hour to allow the turbo pump to pump residual air in the instrument. After 1 hour, valve #1 was opened and mass spectrometry data was recorded.

**Auger Electron Spectroscopy (AES) with a Scanning Auger Nanoprobe**

The AES and elemental mapping measurements were conducted in Stanford Nano Shared Facilities (SNSF) using a PHI 700 Scanning Auger Nanoprobe. After the battery has reached its designated state, it was disassembled inside an argon-filled glovebox. The graphite electrode was then removed from the battery and sealed inside an aluminum laminated pouch (MTI, EQ-alf-100-210). The sample was then transferred to the facility and introduced into the instrument for measurements. During this transfer process, it was important to minimize the sample's exposure to air.

**Acknowledgements**

This work was supported by a Stanford Bits and Watts Fellowship and a Deng Family gift. Part of this work was performed at the Stanford Nano Shared Facilities (SNSF), supported by the National Science Foundation under award ECCS-2026822.

**Competing interests**

The authors declare no competing interests.

# Reference

1. Zhu, G.; Tian, X.; Tai, H.-C.; Li, Y.-Y.; Li, J.; Sun, H.; Liang, P.; Angell, M.; Huang, C.-L.; Ku, C.-S.; Hung, W.-H.; Jiang, S.-K.; Meng, Y.; Chen, H.; Lin, M.-C.; Hwang, B.-J.; Dai, H., Rechargeable Na/Cl2 and Li/Cl2 batteries. *Nature* **2021,** *596* (7873), 525-530.
2. Lin, M.-C.; Gong, M.; Lu, B.; Wu, Y.; Wang, D.-Y.; Guan, M.; Angell, M.; Chen, C.; Yang, J.; Hwang, B.-J.; Dai, H., An ultrafast rechargeable aluminium-ion battery. *Nature* **2015,** *520* (7547), 324-328.
3. Sun, H.; Zhu, G.; Zhu, Y.; Lin, M.-C.; Chen, H.; Li, Y.-Y.; Hung, W. H.; Zhou, B.; Wang, X.; Bai, Y.; Gu, M.; Huang, C.-L.; Tai, H.-C.; Xu, X.; Angell, M.; Shyue, J.-J.; Dai, H., High-Safety and High-Energy-Density Lithium Metal Batteries in a Novel Ionic-Liquid Electrolyte. *Advanced Materials* **2020,** *32* (26), 2001741.
4. Pan, C.-J.; Yuan, C.; Zhu, G.; Zhang, Q.; Huang, C.-J.; Lin, M.-C.; Angell, M.; Hwang, B.-J.; Kaghazchi, P.; Dai, H., An operando X-ray diffraction study of chloroaluminate anion-graphite intercalation in aluminum batteries. *Proceedings of the National Academy of Sciences* **2018,** *115* (22), 5670-5675.
5. Sun, H.; Zhu, G.; Xu, X.; Liao, M.; Li, Y.-Y.; Angell, M.; Gu, M.; Zhu, Y.; Hung, W. H.; Li, J.; Kuang, Y.; Meng, Y.; Lin, M.-C.; Peng, H.; Dai, H., A safe and non-flammable sodium metal battery based on an ionic liquid electrolyte. *Nature Communications* **2019,** *10* (1), 3302.
6. Angell, M.; Zhu, G.; Lin, M.-C.; Rong, Y.; Dai, H., Ionic Liquid Analogs of AlCl3 with Urea Derivatives as Electrolytes for Aluminum Batteries. *Advanced Functional Materials* **2020,** *30* (4), 1901928.
7. Zhu, G.; Angell, M.; Pan, C.-J.; Lin, M.-C.; Chen, H.; Huang, C.-J.; Lin, J.; Achazi, A. J.; Kaghazchi, P.; Hwang, B.-J.; Dai, H., Rechargeable aluminum batteries: effects of cations in ionic liquid electrolytes. *RSC Advances* **2019,** *9* (20), 11322-11330.
8. Liang, P.; Zhang, H.; Su, Y.; Huang, Z.; Wang, C.-A.; Zhong, M., In situ preparation of a binder-free nano-cotton-like CuO–Cu integrated anode on a current collector by laser ablation oxidation for long cycle life Li-ion batteries. *Journal of Materials Chemistry A* **2017,** *5* (37), 19781-19789.
9. Shen, X.; Li, Y.; Qian, T.; Liu, J.; Zhou, J.; Yan, C.; Goodenough, J. B., Lithium anode stable in air for low-cost fabrication of a dendrite-free lithium battery. *Nature Communications* **2019,** *10* (1), 900.
10. Barpanda, P.; Oyama, G.; Nishimura, S.-i.; Chung, S.-C.; Yamada, A., A 3.8-V earth-abundant sodium battery electrode. *Nature Communications* **2014,** *5* (1), 4358.
11. Tsai, C.-Y.; Tai, H.-C.; Su, C.-A.; Chiang, L.-M.; Li, Y.-Y., Activated Microporous Carbon Nanospheres for Use in Supercapacitors. *ACS Applied Nano Materials* **2020,** *3* (10), 10380-10388.
12. Frank, O.; Mohr, M.; Maultzsch, J.; Thomsen, C.; Riaz, I.; Jalil, R.; Novoselov, K. S.; Tsoukleri, G.; Parthenios, J.; Papagelis, K.; Kavan, L.; Galiotis, C., Raman 2D-Band Splitting in Graphene: Theory and Experiment. *ACS Nano* **2011,** *5* (3), 2231-2239.
13. Mohiuddin, T. M. G.; Lombardo, A.; Nair, R. R.; Bonetti, A.; Savini, G.; Jalil, R.; Bonini, N.; Basko, D. M.; Galiotis, C.; Marzari, N.; Novoselov, K. S.; Geim, A. K.; Ferrari, A. C., Uniaxial strain in graphene by Raman spectroscopy: G peak splitting, Gruneisen parameters, and sample orientation. *Physical Review B* **2009,** *79* (20), 205433.
14. Eckmann, A.; Felten, A.; Mishchenko, A.; Britnell, L.; Krupke, R.; Novoselov, K. S.; Casiraghi, C., Probing the Nature of Defects in Graphene by Raman Spectroscopy. *Nano Letters* **2012,** *12* (8), 3925-3930.

15. Malard, L. M.; Pimenta, M. A.; Dresselhaus, G.; Dresselhaus, M. S., Raman spectroscopy in graphene. *Physics Reports* **2009,** *473* (5), 51-87.
16. Ni, Z. H.; Yu, T.; Lu, Y. H.; Wang, Y. Y.; Feng, Y. P.; Shen, Z. X., Uniaxial Strain on Graphene: Raman Spectroscopy Study and Band-Gap Opening. *ACS Nano* **2008,** *2* (11), 2301-2305.
17. Abraham, K. M.; Pitts, L.; Kilroy, W. P., Physical and Chemical Characteristics of Hermetically Sealed High Rate Li / SOCl2 C‐Cells. *Journal of The Electrochemical Society* **1985,** *132* (10), 2301-2308.
18. Spotnitz, R. M.; Yeduvaka, G. S.; Nagasubramanian, G.; Jungst, R., Modeling self-discharge of Li/SOCl2 cells. *Journal of Power Sources* **2006,** *163* (1), 578-583.
19. Abraham, K. M.; Mank, R. M., Some Chemistry in the Li / SOCl2 Cell. *Journal of The Electrochemical Society* **1980,** *127* (10), 2091-2096.
20. Abraham, K. M.; Mank, R. M.; Holleck, G. L. *Investigations of the safety of Li/SOCl2 batteries*; November 01, 1979, 1979.
21. Gao, Y.; Chen, L.; Quan, M.; Zhang, G.; Zheng, Y.; Zhao, J., A series of new Phthalocyanine derivatives with large conjugated system as catalysts for the Li/SOCl2 battery. *Journal of Electroanalytical Chemistry* **2018,** *808*, 8-13.
22. Wang, D.; Jiang, J.; Pan, Z.; Li, Q.; Zhu, J.; Tian, L.; Shen, P. K., The Effects of Pore Size on Electrical Performance in Lithium-Thionyl Chloride Batteries. *Frontiers in Materials* **2019,** *6* (245).
23. 김명수, Electromagnetic Interference Shielding Properties of CO2 Activated Carbon Black Filled Polymer Coating Materials. *Carbon letters* **2008,** *9* (4), 298-302.
24. Xu, Z.; Yan, H.; Yao, K.; Li, K.; Li, J.; Jiang, K.; Li, Z., Enhanced Stable and High Voltage of Li/SOCl2 Battery Catalyzed by FePc Particulates Fixed on Activated Carbon Substrates. *Journal of The Electrochemical Society* **2021,** *168* (10), 100528.
25. Venkatasetty, H. V.; Saathoff, D. J., Properties of LiAlCl4 ‐ SOCl2 Solutions for Li / SOCl2 Battery. *Journal of The Electrochemical Society* **1981,** *128* (4), 773-777.
26. Yang, G.; Li, Y.; Liu, S.; Zhang, S.; Wang, Z.; Chen, L., LiFSI to improve lithium deposition in carbonate electrolyte. *Energy Storage Materials* **2019,** *23*, 350-357.
27. Heist, A.; Lee, S.-H., Improved Stability and Rate Capability of Ionic Liquid Electrolyte with High Concentration of LiFSI. *Journal of The Electrochemical Society* **2019,** *166* (10), A1860-A1866.
28. Liu, X.; Shen, C.; Gao, N.; Hou, Q.; Song, F.; Tian, X.; He, Y.; Huang, J.; Fang, Z.; Xie, K., Concentrated electrolytes based on dual salts of LiFSI and LiODFB for lithium-metal battery. *Electrochimica Acta* **2018,** *289*, 422-427.
29. Bhatt, A. I.; Best, A. S.; Huang, J.; Hollenkamp, A. F., Application of the N-propyl-N-methyl-pyrrolidinium Bis(fluorosulfonyl)imide RTIL Containing Lithium Bis(fluorosulfonyl)imide in Ionic Liquid Based Lithium Batteries. *Journal of The Electrochemical Society* **2010,** *157* (1), A66.
30. Wassei, J. K.; Cha, K. C.; Tung, V. C.; Yang, Y.; Kaner, R. B., The effects of thionyl chloride on the properties of graphene and graphene–carbon nanotube composites. *Journal of Materials Chemistry* **2011,** *21* (10), 3391-3396.
31. Chacón-Torres, J. C.; Wirtz, L.; Pichler, T., Raman spectroscopy of graphite intercalation compounds: Charge transfer, strain, and electron–phonon coupling in graphene layers. *physica status solidi (b)* **2014,** *251* (12), 2337-2355.

32. Balabajew, M.; Reinhardt, H.; Bock, N.; Duchardt, M.; Kachel, S.; Hampp, N.; Roling, B., In-Situ Raman Study of the Intercalation of Bis(trifluoromethylsulfonyl)imid Ions into Graphite inside a Dual-Ion Cell. *Electrochimica Acta* **2016,** *211*, 679-688.
33. Mohandas, K. S.; Sanil, N.; Noel, M.; Rodriguez, P., Electrochemical intercalation of aluminium chloride in graphite in the molten sodium chloroaluminate medium. *Carbon* **2003,** *41* (5), 927-932.
34. Sole, C.; Drewett, N. E.; Hardwick, L. J., In situ Raman study of lithium-ion intercalation into microcrystalline graphite. *Faraday Discussions* **2014,** *172* (0), 223-237.
35. Cançado, L. G.; Takai, K.; Enoki, T.; Endo, M.; Kim, Y. A.; Mizusaki, H.; Jorio, A.; Coelho, L. N.; Magalhães-Paniago, R.; Pimenta, M. A., General equation for the determination of the crystallite size La of nanographite by Raman spectroscopy. *Applied Physics Letters* **2006,** *88* (16), 163106.
36. Pimenta, M. A.; Dresselhaus, G.; Dresselhaus, M. S.; Cançado, L. G.; Jorio, A.; Saito, R., Studying disorder in graphite-based systems by Raman spectroscopy. *Physical Chemistry Chemical Physics* **2007,** *9* (11), 1276-1290.
37. Das, A.; Chakraborty, B.; Sood, A. K., Raman spectroscopy of graphene on different substrates and influence of defects. *Bulletin of Materials Science* **2008,** *31* (3), 579-584.
38. Gayathri, S.; Jayabal, P.; Kottaisamy, M.; Ramakrishnan, V., Synthesis of few layer graphene by direct exfoliation of graphite and a Raman spectroscopic study. *AIP Advances* **2014,** *4* (2), 027116.
39. Lee, S.-M.; Lee, S.-H.; Roh, J.-S., Analysis of Activation Process of Carbon Black Based on Structural Parameters Obtained by XRD Analysis. *Crystals* **2021,** *11* (2), 153.
40. Stobinski, L.; Lesiak, B.; Malolepszy, A.; Mazurkiewicz, M.; Mierzwa, B.; Zemek, J.; Jiricek, P.; Bieloshapka, I., Graphene oxide and reduced graphene oxide studied by the XRD, TEM and electron spectroscopy methods. *Journal of Electron Spectroscopy and Related Phenomena* **2014,** *195*, 145-154.
41. Kumar, V.; Kumar, A.; Lee, D.-J.; Park, S.-S., Estimation of Number of Graphene Layers Using Different Methods: A Focused Review. *Materials* **2021,** *14* (16), 4590.
42. Asenbauer, J.; Eisenmann, T.; Kuenzel, M.; Kazzazi, A.; Chen, Z.; Bresser, D., The success story of graphite as a lithium-ion anode material – fundamentals, remaining challenges, and recent developments including silicon (oxide) composites. *Sustainable Energy & Fuels* **2020,** *4* (11), 5387-5416.

# Supporting Information

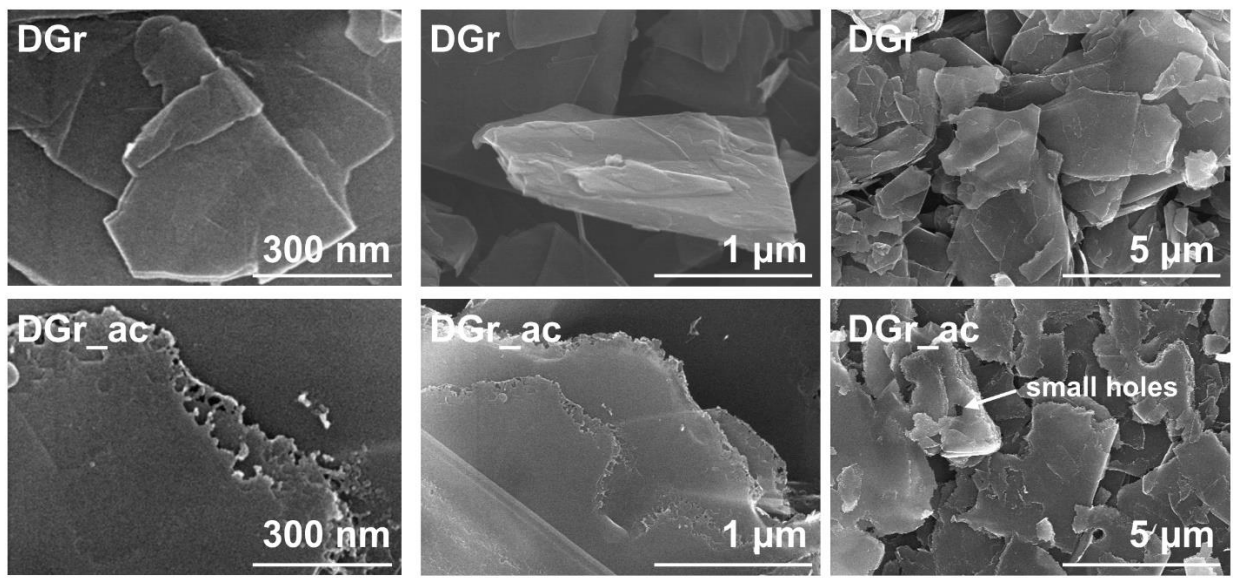

**Figure S1. SEM images of the as-received DGr and the activated DGr_ac. Top row:** SEM images of as-received DGr at different magnifications. **Bottom row:** SEM images of activated DGr_ac at different magnifications. Most of the defects formed after annealing in a $CO_2$ environment at 1000 ºC were at the edge of the graphite flakes, but some small holes were also formed in the middle of the graphite plates (indicated by arrow).

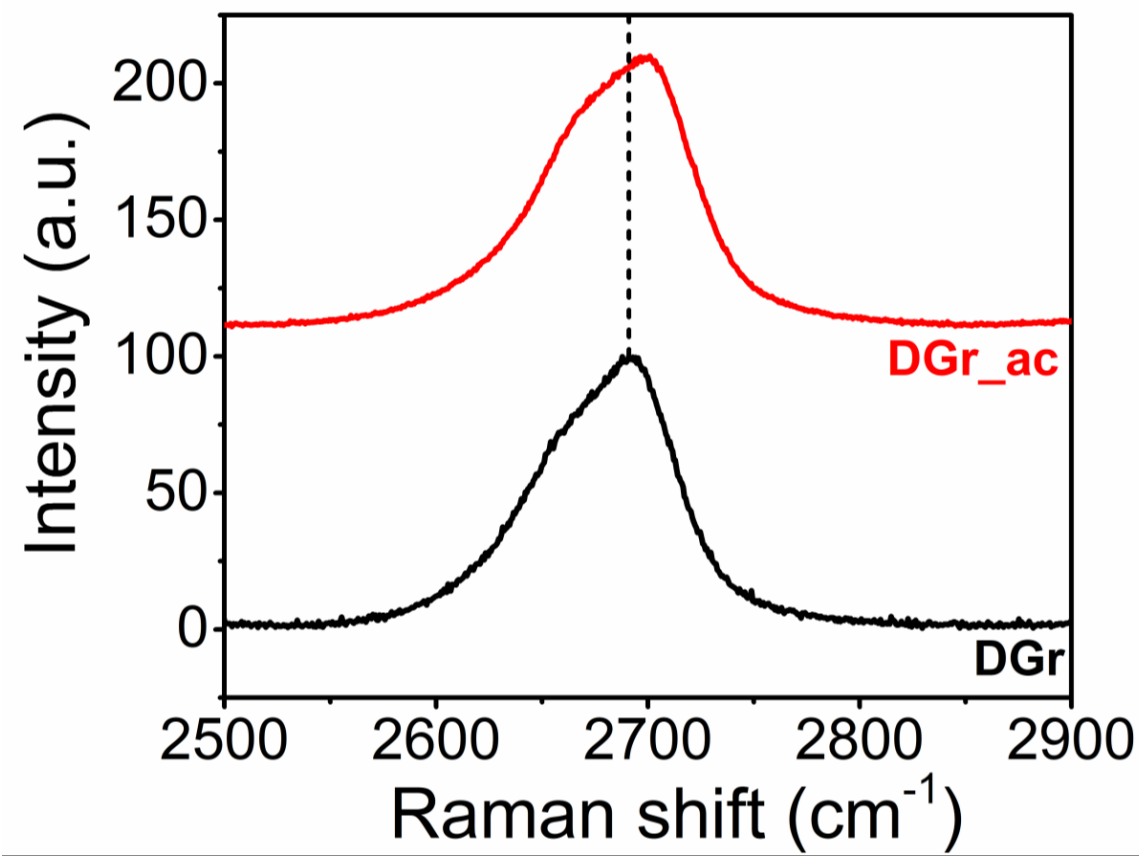

**Figure S2. Raman spectrum in the 2D band region from 2500 cm$^{-1}$ to 2900 cm$^{-1}$ of the as-received DGr and the activated DGr_ac.** A blue shift in the 2D band wavenumber was observed in the activated DGr_ac when compared to the as-received DGr, indicating a decrease in the strain within the activated DGr_ac.

|         | Surface area (m$^2$ g$^{-1}$) | Pore volume (cm$^3$ g$^{-1}$) | Micropore volume (cm$^3$ g$^{-1}$) |
|---------|---|---|---|
| DGr     | 13.11 | 0.05 | 0.00051 (1.02%) |
| DGr_ac  | 18.73 | 0.07 | 0.00220 (3.14%) |

**Table S1. Surface area and pore volume of the as-received DGr and the activated DGr_ac.** After annealing in $CO_2$, DGr_ac had both its surface area and pore volume increased by ~ 42.9% and ~ 40.0%, respectively.

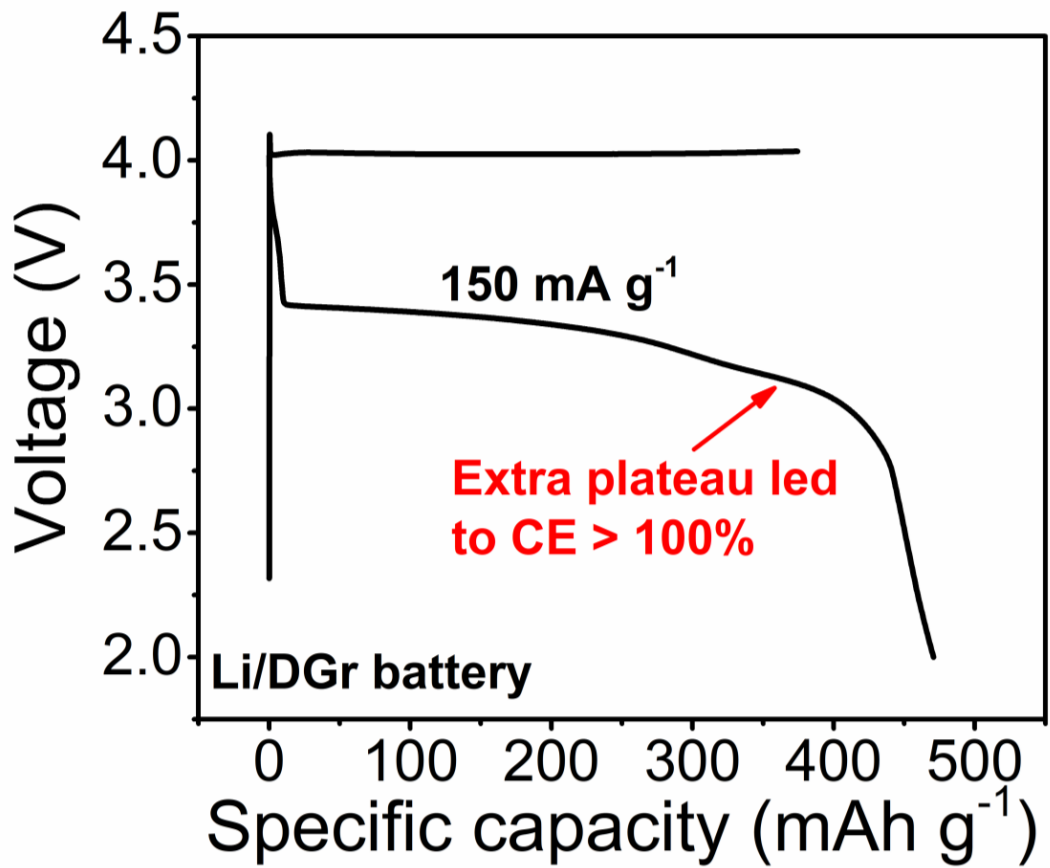

**Figure S3. Typical charge-discharge curve of a Li/DGr battery at 375 mAh g$^{-1}$ when the coulombic efficiency was higher than 100%.** The extra plateau towards the end of discharging at ~ 3.15 V was likely due to extra SOCl$_2$ reduction caused by the 'in-situ activation' of DGr over battery cycling. The loading of DGr was ~ 4.3 mg cm$^{-2}$. The charging step was controlled by setting the charging time to be 2.5 hours at 150 mA g$^{-1}$ current (charging capacity = 375 mAh g$^{-1}$). The discharging step was controlled by setting a discharge cutoff voltage of 2 V.

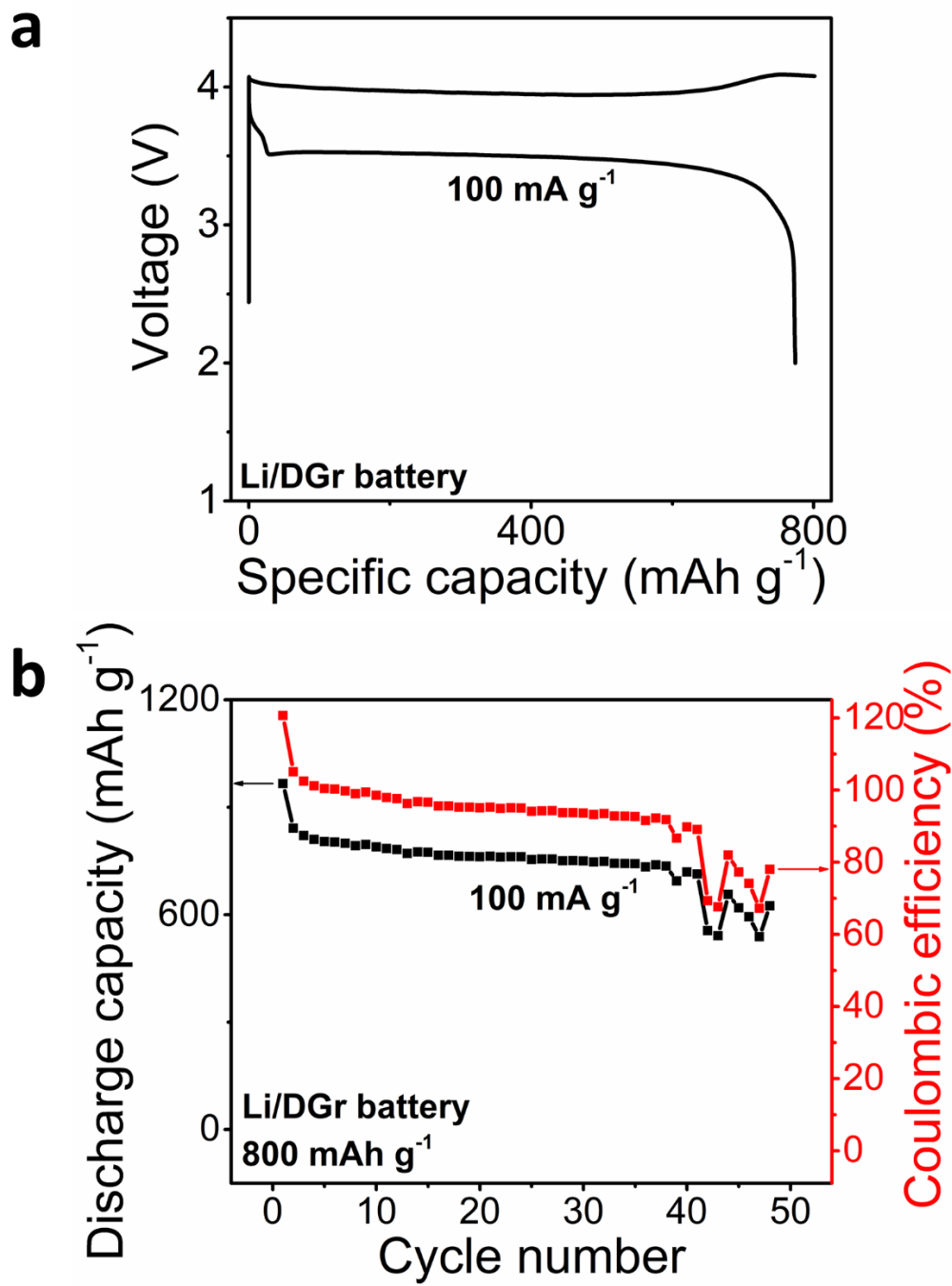

**Figure S4. Li/DGr battery cycling performance at 800 mAh g$^{-1}$. a,** Typical charge-discharge curve of a Li/DGr battery at 800 mAh g$^{-1}$. The shape of the curve was similar to that at 375 mAh g$^{-1}$, with the main charging and discharging plateaus simply extended. **b,** Cycling performance of a Li/DGr battery at 800 mAh g$^{-1}$ with 100 mA g$^{-1}$ current. The battery displayed an inferior cycle life of less than 50 cycles. The loading of DGr was ~ 4.5 mg cm$^{-2}$. The charging step was controlled by setting the charging time to be 8 hours at 100 mA g$^{-1}$ current (charging capacity = 800 mAh g$^{-1}$). The discharging step was controlled by setting a discharge cutoff voltage of 2 V.

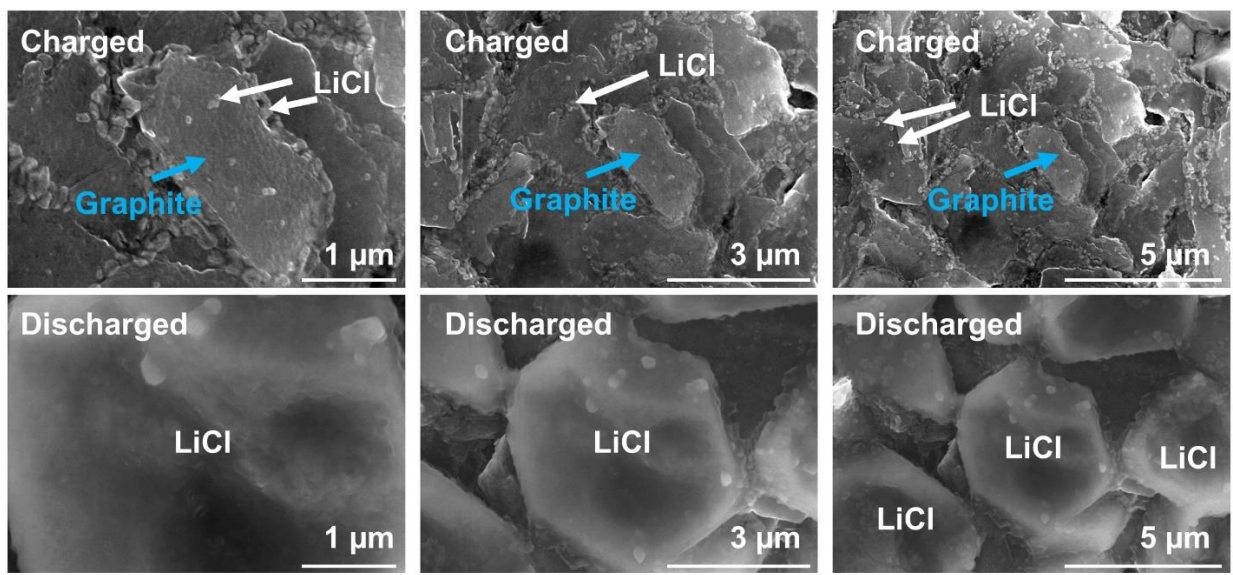

**Figure S5. SEM images of DGr_ac electrodes after charging and discharging to 800 mAh g$^{-1}$. Top row:** SEM images of DGr_ac electrode after charging to 800 mAh g$^{-1}$. Most of the LiCl on the electrode were removed/oxidized to form Cl$_2$, and the graphite flakes underneath were easily detected. **Bottom row:** SEM images of DGr_ac electrode after discharging to 800 mAh g$^{-1}$. A very thick layer of LiCl was formed on the DGr_ac electrode, due to the reduction of Cl$_2$ to LiCl. No graphite flakes were detected as they were covered by LiCl crystals.

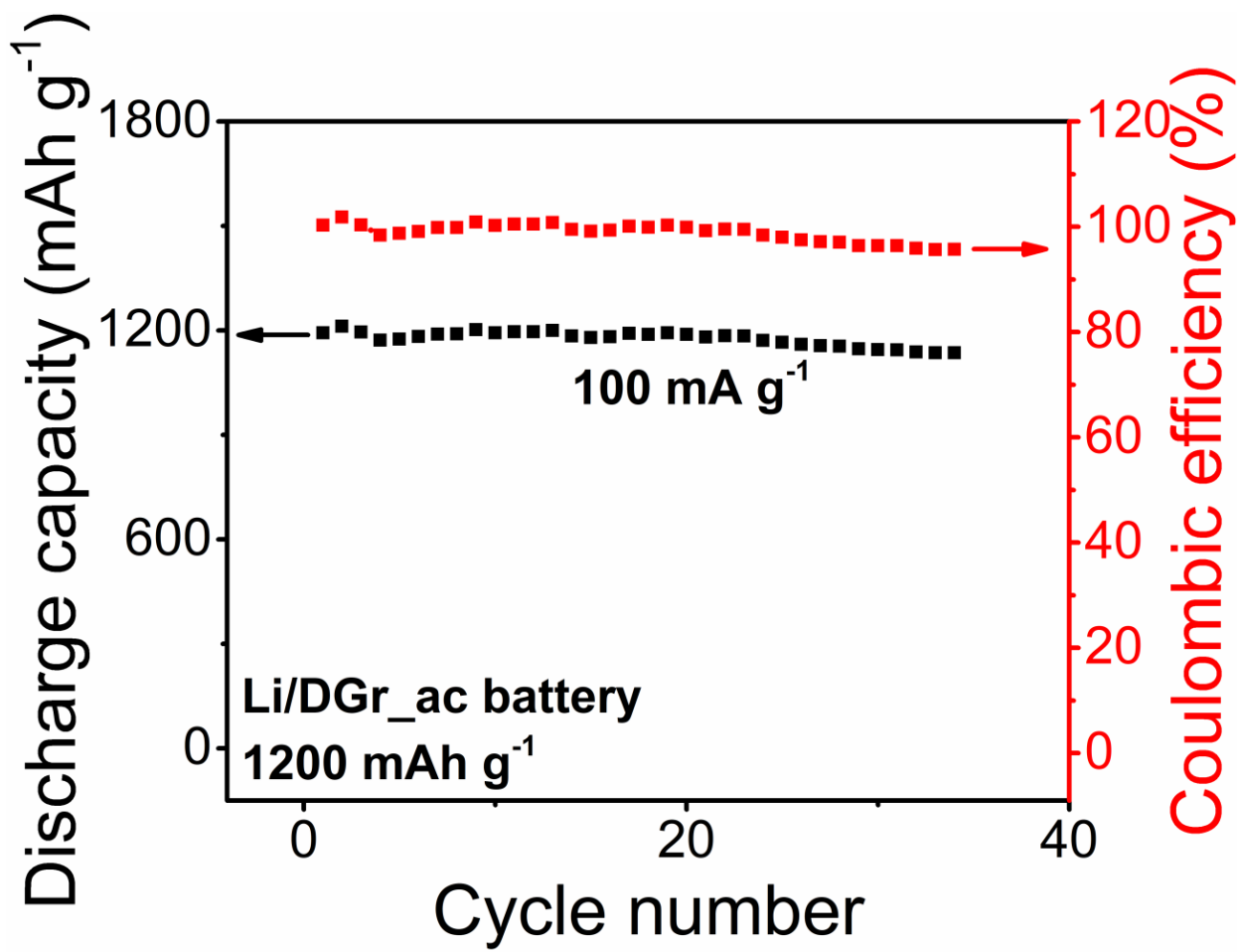

**Figure S6. Cycling performance of a Li/DGr_ac battery at 1200 mAh g$^{-1}$ with 100 mA g$^{-1}$ current.** The battery was cyclable at 1200 mAh g$^{-1}$ for more than 35 cycles. The loading of DGr_ac was ~ 3.0 mg cm$^{-2}$. The charging step was controlled by setting the charging time to be 12 hours at 100 mA g$^{-1}$ current (charging capacity = 1200 mAh g$^{-1}$). The discharging step was controlled by setting a discharge cutoff voltage of 2 V.

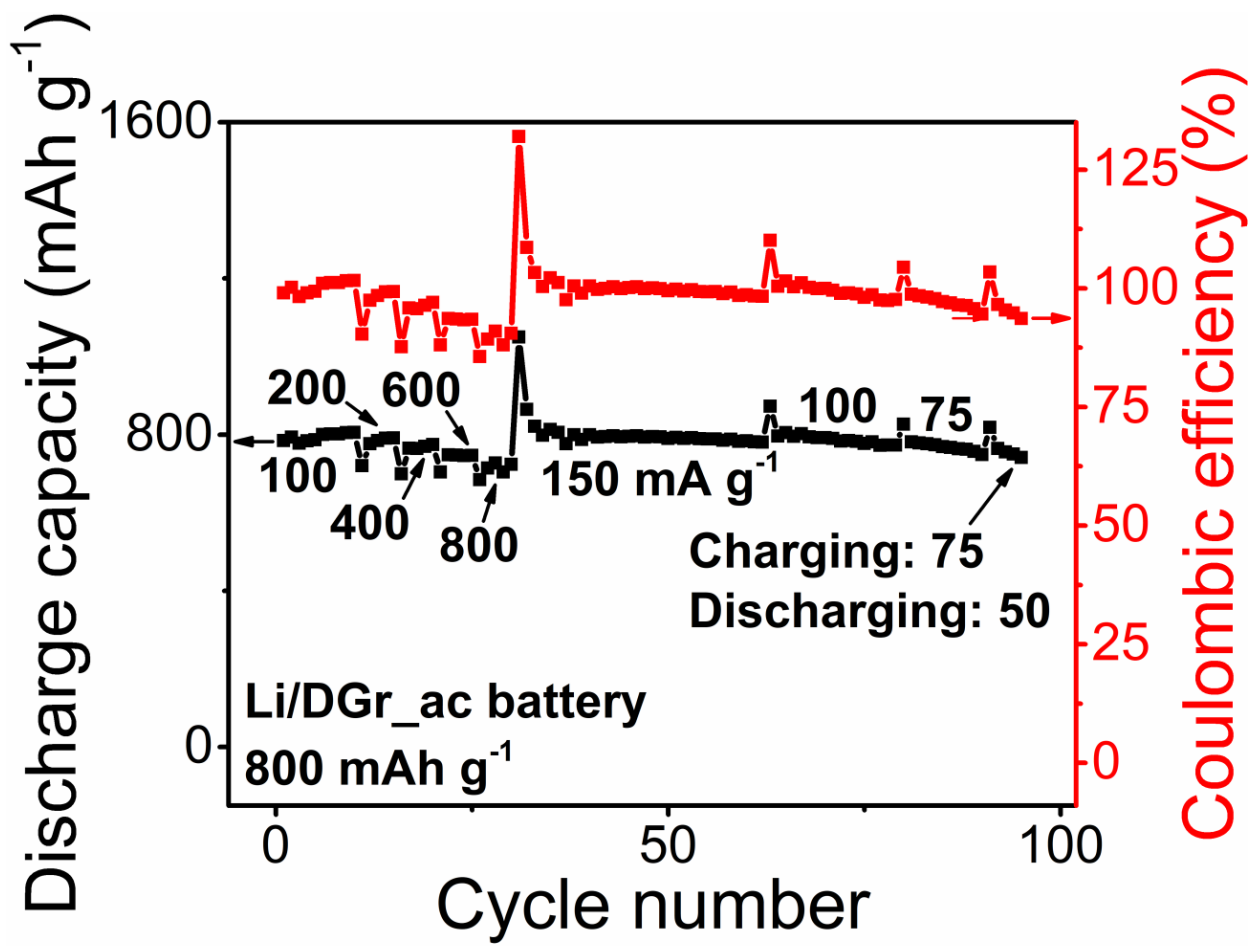

**Figure S7. Cycling performance of a Li/DGr_ac battery at 800 mAh g$^{-1}$ with various currents (50 mA g$^{-1}$, 75 mA g$^{-1}$, 100 mA g$^{-1}$, 150 mA g$^{-1}$, 200 mA g$^{-1}$, 400 mA g$^{-1}$, 600 mA g$^{-1}$, and 800 mA g$^{-1}$).** The battery was able to cycle under all these current conditions and with its coulombic efficiency slightly decreased as the current became higher. The charging step was controlled by setting the charging time depending on the current condition. The discharging step was controlled by setting a discharge cutoff voltage of 2 V.

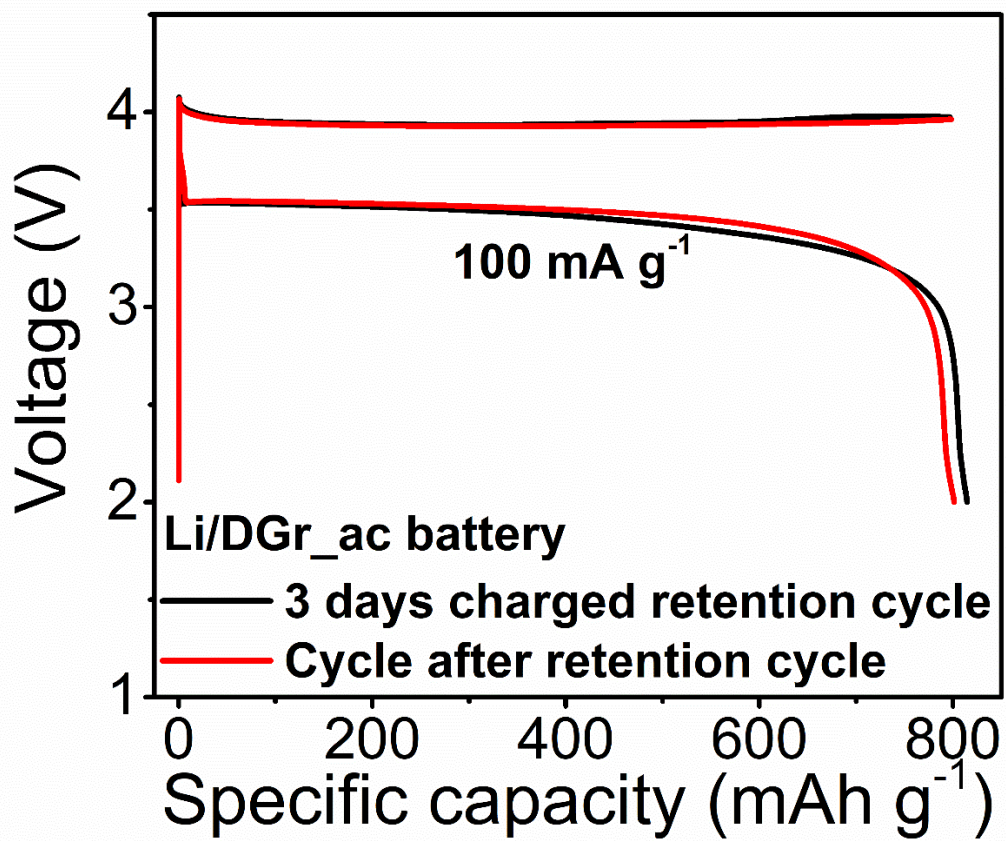

**Figure S8. Charge-discharge curves of Li/DGr_ac batteries in a 3-day charged retention cycle and the cycle immediately after that retention cycle. Black curve**: charge-discharge curve after holding the battery in charged state for 3 days. The main discharging plateau at ~ 3.54 V was shortened and a lower discharging plateau at ~ 3.35 V was observed. **Red curve**: charge-discharge curve after the 3-day charged retention cycle. The normal charge-discharge behavior was immediately restored.

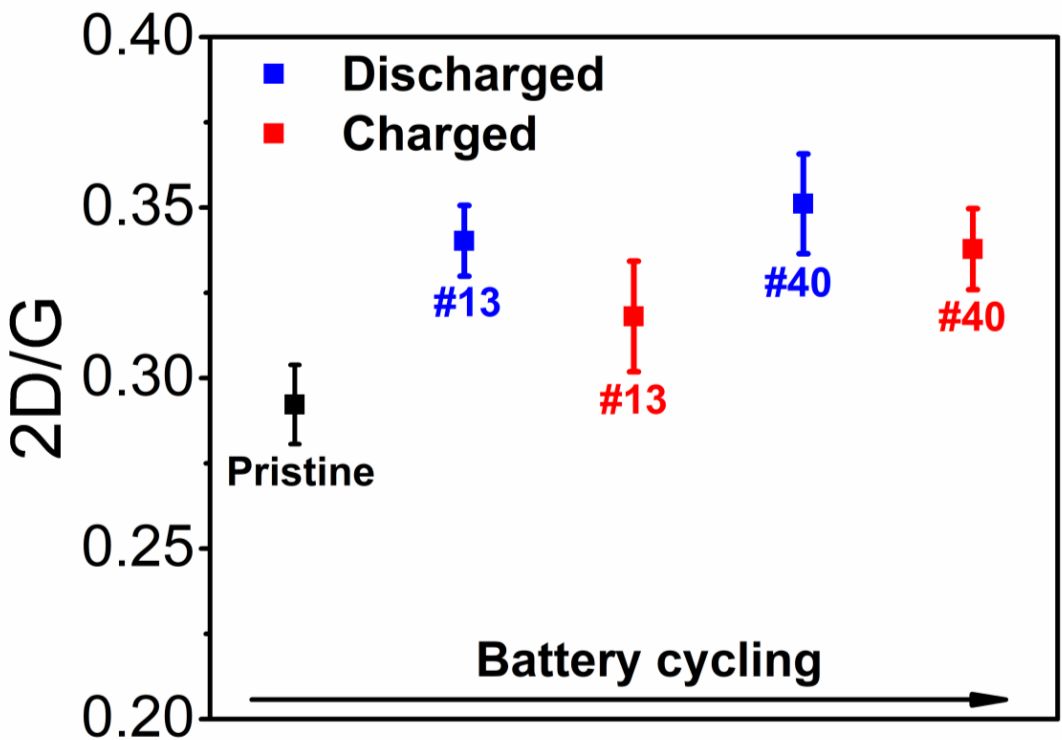

**Figure S9. The intensity ratio of the 2D Raman band (~ 2700 cm$^{-1}$) to the G Raman band (~ 1570 cm$^{-1}$) of DGr_ac in different states of battery cycling after washing using DIUF water.** As the battery cycled more, this ratio of 2D/G increased, which corresponded to a decrease in the number of graphene layers in the graphite flake or exfoliations.

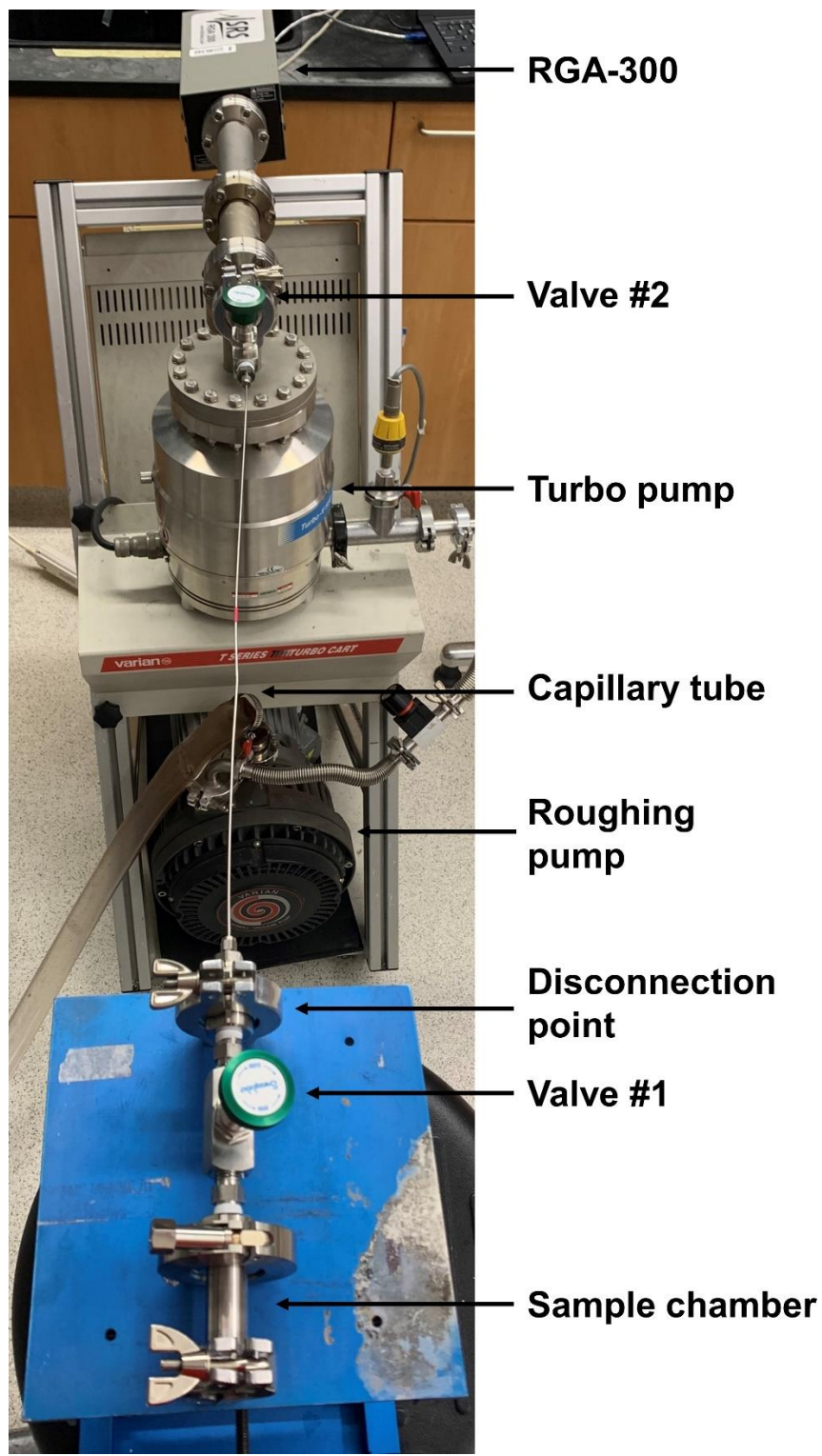

**Figure S10. Setup of the RGA-300 instrument for mass spectrometry measurement.** See Methods for details.

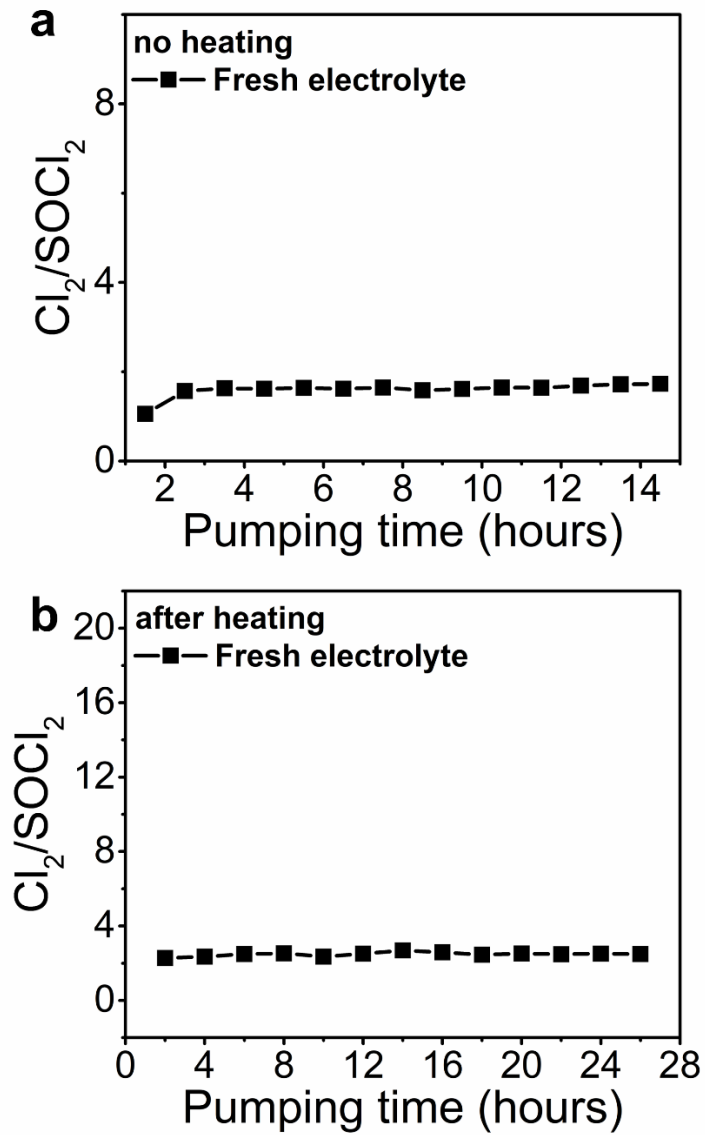

**Figure S11. Mass spectrometry studies of fresh electrolyte before (a) and after (b) heating at 80 ºC for 2 hours.** The ratio between the detected $Cl_2$ pressure to the detected $SOCl_2$ pressure remained nearly constant throughout the entire experiment.